\documentclass[prd,twocolumn]{revtex4}
\usepackage{graphicx}
\usepackage{amssymb}

\begin{document}

\title{{\bf Loop Quantum Gravity and Ultra High Energy Cosmic Rays}}

\author{Jorge Alfaro and  Gonzalo Palma}

\affiliation{\linespread{1} Facultad de F\'{\i}sica, Pontificia Universidad Cat\'{o}lica de Chile \\
Casilla 306, Santiago 22, Chile.
\\ {\tt jalfaro@puc.cl, \  gpalma@astro.puc.cl}}

\date{\today}


\begin{abstract}
There are two main sets of data for the observed spectrum of ultra
high energy cosmic rays (those cosmic rays with energies greater
than $\sim 4 \times 10^{18}$ eV), the High Resolution Fly's Eye
(HiRes) collaboration group observations, which seem to be
consistent with the predicted theoretical spectrum (and therefore
with the theoretical limit known as the Greisen-Zatsepin-Kuzmin
cutoff), and the observations from the Akeno Giant Air Shower
Array (AGASA) collaboration group, which reveal an abundant flux
of incoming particles with energies above $1 \times 10^{20}$ eV
violating the Greisen-Zatsepin-Kuzmin cutoff. As an explanation of
this anomaly it has been suggested that quantum-gravitational
effects may be playing a decisive role in the propagation of ultra
high energy cosmic rays. In this article we take the loop quantum
gravity approach. We shall provide some techniques to
establish and analyze new constraints on the loop quantum gravity
parameters arising from both sets of data, HiRes and AGASA . We
shall also study their effects on the predicted spectrum for ultra
high energy cosmic rays. As a result we will state the possibility
of reconciling the AGASA observations.
\end{abstract}

\maketitle


\section{Introduction} \label{INTRO}

In this article we are concerned with the observation of ultra
high energy cosmic rays (UHECR), i.e. those cosmic rays with
energies greater than $\sim 4 \times 10^{18}$ eV. Although not
completely clear, it has been suggested that these high energy
particles are possibly heavy nuclei \cite{exp1, Ave} (we will
assume here that they are protons) and, by virtue of the isotropic
distribution with which they arrive to us, that they originate in
extragalactic sources.

A detailed understanding of the origin and nature of UHECR is far
from being achieved; the way in which the observed cosmic ray
spectrum appears to us still is a mystery and a matter of great
debate. A first subject of interest (faced with the lack of
reasonable mechanisms) is how such energetic particles have been
accelerated to energies well above $4 \times 10^{18}$ eV by its
sources. A second subject of interest is the study of their
propagation in open space through the cosmic microwave background
radiation (CMBR), whose presence necessarily produce a friction on
UHECR making them release energy in the form of secondary
particles and affecting their possibility to reach great
distances. A first estimation of the characteristic distance that
UHECR can reach before losing most of their energy was
simultaneously made in 1966 by K. Greisen \cite{Greisen} and G. T.
Zatsepin \& V. A. Kuzmin \cite{Zatsepin & Kuzmin}, who showed that
the observation of cosmic rays with energies greater than $4
\times 10^{19}$ eV should be greatly suppressed. This energy ($4
\times 10^{19}$ eV) is usually referred as to the GZK cutoff
energy. Similarly and a few years later, F.W. Stecker
\cite{Stecker} calculated the mean life time for protons as a
function of their energy, giving a more accurate perspective of
the energy behavior of the cutoff and showing that cosmic rays
with energies above $1 \times 10^{20}$ eV should not travel more
than $\sim 100$ Mpc. More detailed approaches to the GZK-cutoff
feature have been made since these first estimations. For example
V. Berezinsky \& S.I. Grigorieva \cite{Berezinsky2}, V. Berezinsky
{\it et al.} \cite{Berezinsky} and S.T. Scully \& F.W. Stecker
\cite{Scully & Stecker} have made progress in the theoretical
study of the spectrum $J(E)$ (i.e. the flux of arriving particles
as a function of the observed energy $E$) that UHECR should
present. As a result, the GZK cutoff exists in the form of a
suppression in the predicted flux of cosmic rays with energies
above $\sim 8 \times 10^{19}$ eV.

At present there are two main different sets of data for the
observed flux $J(E)$ in its most energetic sector ($E > 4 \times
10^{18}$ eV). On one hand we have the observations from the High
Resolution Fly's Eye (HiRes) collaboration group \cite{HiRes},
which seem to be consistent with the predicted theoretical
spectrum and, therefore, with the presence of the GZK cutoff.
Meanwhile, on the other hand, we have the observations from the
Akeno Giant Air Shower Array (AGASA) collaboration group
\cite{exp3}, which reveal an abundant flux of incoming cosmic rays
with energies above $1 \times 10^{20}$ eV. The appearance of these
high energy events is greatly opposed to the predicted GZK cutoff,
and a great challenge that has motivated a vast amount of new
ideas and mechanisms to explain this phenomenon \cite{Tkachev 1,
Tkachev 2, Tkachev 4, Monopoles, Z-Burst1, Z-Burst2, Z-Burst3,
Super-relic}. If the AGASA observations are correct then, since
there are no known active objects in our neighborhood (let us say
within a radius $R \simeq 100$ Mpc) able to act as sources of such
energetic particles and since their arrival is mostly isotropic
(without any privileged local source), we are forced to conclude
that these cosmic rays come from distances larger than $100$ Mpc.
This is commonly referred as the Greisen-Zatsepin-Kuzmin (GZK)
anomaly.

One of the interesting notions emerging from the possible
existence of the GZK anomaly is that, since ultra high energy
cosmic rays involve the highest energy events registered up to
now, then  a possible framework to understand and explain
this phenomena could be of a quantum-gravitational nature
\cite{KIFUNE, Amelino-Piran1, Ellis, Amelino-Piran2, Amelino,
Alfaro & Palma}. This possibility is indeed very exciting if we
consider the present lack of empirical support for the different
approaches to the problem of gravity quantization. In the context
of the UHECR phenomena, all these different approaches motivated
by different quantum gravity formulations, have usually converged
on a common path to solve and explain the GZK anomaly: the
introduction of effective models for the description of high
energy particle propagation. These effective models, pictures of
the yet unknown full quantum gravity theory, offer the possibility
to modify conventional physics through new terms in the equations
of motion (now effective equations of motion), leading to the
eventual breakup of fundamental symmetries such as Lorentz
invariance (expected to be preserved at the fundamental level).
These Lorentz symmetry breaking mechanisms are usually referred as
Lorentz invariance violations (LIV's) if the break introduce a
privileged reference frame, or Lorentz invariance deformations
(LID's), if such a reference frame is absent \cite{Amelino2,
Magueijo+Smolin, Kowalski-Glikman}. Its appearance on theoretical
as well as phenomenological grounds (such as high energy
astrophysical phenomena) has been widely studied, and offers a
large and rich amount of new signatures that deserve attention
\cite{colladay1, colladay2, Amelino & Ellis, bertolami1,
bertolami2, Liberati1, Major, Noncommutativity, Liberati2}.

To deepen the above ideas, we have adopted the loop quantum
gravity (LQG) theory \cite{LQG1, LQG2}, one of the proposed
alternatives for the yet nonexistent theory of quantum gravity. It
is possible to study LQG through effective theories that take into
consideration matter-gravity couplings. In this line, in the works
of J. Alfaro {\it et al}. \cite{Neutrinos, Photons, AMU}, the
effects of the loop structure of space at the Planck level are
treated semiclassically through a coarse-grained approximation. An
interesting feature of these methods is the explicit appearance of
the Plank scale $l_{p}$ and the appearance of a new length scale
$\mathcal{L} \gg l_{p}$ (called the ``weave'' scale), such that
for distances $d \ll \mathcal{L}$ the quantum loop structure of
space is manifest, while for distances $d \geqslant \mathcal{L}$
the continuous flat geometry is regained. The presence of these
two scales in the effective theories has the consequence of
introducing LIV's to the dispersion relations $E=E(p)$ for
particles with energy $E$ and momentum $p$. It can be shown that
these LIV's can significantly modify the kinematical conditions
for a reaction to take place. For instance, as shown in detail in
\cite{Coleman & Glashow}, if the dispersion relation for a
particle $i$ is (from here on, $\hbar = c = 1$)
\begin{eqnarray}
E_{i}^{2} = A_{i}^{2} p_{i}^{2} + m_{i}^{2}
\end{eqnarray}
(where $E_{i}$, $p_{i}$ and $m_{i}$ are the respective energy,
momentum and mass of the $i$th particle, and $A_{i}$ is a LIV
parameter that can be interpreted as the maximum velocity of the
$i$th particle) then the threshold condition for a reaction to
take place can be substantially modified if the difference $\delta
A = A_{a} - A_{b}$ is non zero ($a$ and $b$ are two particles
involved in the reaction leading to the mentioned threshold)
\cite{Coleman & Glashow}. An interesting consequence of the above
situation ---for the UHECR phenomenology--- is that the
kinematical conditions for a reaction between a primary cosmic ray
and a CMBR photon can be modified, leading to new effects and
predictions such as an abundant flux of cosmic rays well beyond
the GZK cutoff energy (explaining in this way the AGASA
observations).

 The purpose of this paper is to provide some techniques to
establish and analyze new constraints on the LQG parameters (or
any other LIV parameters), that will confidently rise when the
experimental situation is clarified in a reliably way up to a
certain energy scale. In the present case, and for the practical
purpose of this paper, we shall assume that such energy scale is
currently $4 \times 10^{19}$ eV. Also, we shall attempt to
predict (under certain assumptions) a modified UHECR spectrum
arising from the LQG corrections to the conventional theory, and
consistent with the AGASA observations  (though we shall
analyze both, HiRes and AGASA sets of data, throughout this paper
we will be more concerned with the possibility that the AGASA
results are the correct one). To accomplish these goals, we have
organized this article as follows: In section \ref{UHECR}, ``Ultra
high energy cosmic rays'', we shall give a brief self contained
derivation of the conventional spectrum and briefly analyze it
jointly with HiRes and AGASA observations. In section \ref{LQG},
``Loop quantum gravity'', we will present a short outline of loop
quantum gravity and its effective description of fermion and
electromagnetic fields (relevant for the description of UHECR
propagation). In section \ref{Trh}, ``Threshold conditions'', we
will analyze the effects of LQG corrections on the threshold
conditions for the main reactions involved in the UHECR phenomena
to take place. In section \ref{MOD-SPEC}, ``Modified spectrum'',
we will show how the modified kinematics can be relevant to the
theoretical spectrum $J(E)$ of cosmic rays (we will present the
obtained modified spectrum). Section \ref{CONC}, ``Conclusions'',
will be reserved for some final remarks.


\section{Ultra High Energy Cosmic Rays} \label{UHECR}

In this section we will review the main steps in the derivation of
the UHECR spectrum. This presentation will be useful and relevant
for the description of the kinematical effects that LQG
corrections can have on the predicted flux of cosmic rays. The
following material is mainly contained in the works of F.W.
Stecker \cite{Stecker}, Berezinsky {\it et al.} \cite{Berezinsky}
and S.T. Scully \& F.W. Stecker \cite{Scully & Stecker}.

\subsection{General Description}

Two simple and commonly used assumptions for the development of
the cosmic ray spectrum are: 1) that the sources are uniformly
distributed in the Universe, and 2) that the generation flux
$F(E_{g})$ of emitted cosmic rays from the sources is correctly
described by a power law behavior of the form $F(E_{g}) \propto
E_{g}^{-\gamma_{g}}$, where $E_{g}$ is the energy of the emitted
particle and  $\gamma_{g}$ is the generation index.

One of the main quantities in the calculation of the UHECR
spectrum is the energy loss $- E^{-1} dE/dt$. This quantity
describes the rate at which a cosmic ray loses energy, and takes
into consideration two chief contributions: the energy loss due to
the redshift attenuation and the energy loss due to collisions
with the CMBR photons. This last contribution depends, at the same
time, on the cross sections $\sigma$ and the inelasticities $K$ of
the interactions produced during the propagation of protons in the
extragalactic medium, as well as on the CMBR spectrum. The most
important reactions taking place in the description of proton's
propagation (and which produce the release of energy in the form
of particles) are the pair creation
\begin{equation}
p + \gamma \rightarrow p + e^{-} + e^{+}, \label{pair}
\end{equation}
and the photo-pion production
\begin{equation}
p + \gamma \rightarrow p + \pi. \label{pion}
\end{equation}
This last reaction happens through several channels (for example
the baryonic $\Delta$ and $N$, and mesonic $\rho$ and $\omega$
resonance channels, just to mention some of them) and is the main
reason for the appearance of the GZK cutoff.

\subsection{Some Kinematics}

To study the interaction between protons and the CMBR, it is
useful to distinguish between three reference systems; the
laboratory system $\mathcal{K}$ (which we identify with the
Friedman Robertson Walker (FRW) co-moving reference system), the
center of mass (c.m.) system $\mathcal{K}^{*}$, and the system
where the proton is at rest $\mathcal{K}'$. In terms of these
systems, the photon energy will be expressed as $\omega$ in
$\mathcal{K}$ and as $\epsilon$ in $\mathcal{K}'$. The relation
between both quantities is simply
\begin{eqnarray} \label{eq2: epsilon'}
\epsilon = \gamma \omega (1-\beta \cos \theta),
\end{eqnarray}
where $\gamma = E/m_{p}$ is the Lorentz factor relating
$\mathcal{K}$ and $\mathcal{K}'$, $E$ and $m_{p}$ are the energy
and mass of the incident proton, $\beta = \sqrt{1-\gamma^{-2}}$,
and $\theta$ is the angle between the momenta of the photon and
the proton measured in the laboratory system $\mathcal{K}$.

To determine the total energy $E^{*}_{\mathrm{tot}} = E^{*} +
\epsilon^{*}$ in the c.m. system, it is enough to use the
invariant energy squared $s \equiv
E_{\mathrm{tot}}^{2}-p_{\mathrm{tot}}^{2}$ (where
$E_{\mathrm{tot}} = E + \omega$ and $p_{\mathrm{tot}}$ are the
total energy and momentum in the laboratory system). In this way,
we have
\begin{eqnarray} \label{eq2: E*-s}
E^{* \, 2}_{\mathrm{tot}} = s = m_{p}^{2} + 2 m_{p} \epsilon.
\end{eqnarray}
As a consequence, the Lorentz factor $\gamma_{c}$ which relates
the $\mathcal{K}$ reference system with the $\mathcal{K}^{*}$
system, is
\begin{eqnarray}
\gamma_{c}=\frac{E+\omega}{\sqrt{s}} \simeq \frac{E}{(m_{p}^{2} +
2 m_{p} \epsilon)^{1/2}}.
\end{eqnarray}

Let us consider the relevant case in which the reaction between
the proton and the CMBR photon is of the type
\begin{eqnarray}
p + \gamma \rightarrow a + b,
\end{eqnarray}
where $a$ and $b$ are two final particles of the collision. The
final energies of these particles are easily determined by the
conservation of energy-momentum. In the $\mathcal{K}^{*}$ system
these are
\begin{eqnarray}
E_{a, \, b}^{*}= \frac{1}{2 \sqrt{s}} (s + m_{a, \, b}^{2} - m_{b,
\, a}^{2}).
\end{eqnarray}
Transforming this quantity to the laboratory system, and averaging
with respect to the angle between the directions of the final
momenta, it is possible to find that the final average energy of
$a$ (or $b$) in the laboratory system is
\begin{eqnarray} \label{eq2: E a-b}
<E_{a, \, b}> = \frac{E}{2} \left( 1+ \frac{m_{a, \, b}^{2} -
m_{b, \, a}^{2}}{s} \right).
\end{eqnarray}

The inelasticity $K$ of the reaction is defined as the average
fractional difference $K = \Delta E / E$, where $\Delta E = E -
E_{\mathrm{f}}$ is the difference between the initial energy $E$
and final energy $E_{\mathrm{f}}$ of the proton (in a single
collision with the CMBR photons). For the particular case of the
emission of an arbitrary particle $a$ (that is to say $p + \gamma
\rightarrow p + a$), expression (\ref{eq2: E a-b}) allows us to
write
\begin{eqnarray} \label{eq2: K}
K_{a}(s) = \frac{1}{2} \left( 1 + \frac{m_{a}^{2}-m_{p}^{2}}{s}
\right),
\end{eqnarray}
where $K_{a}$ is the inelasticity of the described process. This
is one of the main quantities involved in the study of the UHECR
spectrum, in particular, when the emitted particle $a$ is a pion.

\subsection{Mean Life $\tau (E)$}

To derive the UHECR spectrum it is imperative to know the mean
life $\tau (E)$ of the cosmic ray (or proton) with energy $E$
propagating in space, due to the attenuation of its energy by the
interactions with the CMBR photons. The mean life $\tau (E)$ is
defined through the relation
\begin{eqnarray} \label{eq2: tau}
\tau (E)^{-1} = \left( -\frac{1}{E} \frac{dE}{dt}
\right)_{\mathrm{col}},
\end{eqnarray}
where the label ``col'' refers to the fact that the energy loss is
due to the collisions with the CMBR photons. To explicitly
determine the form of $\tau (E)$, let us express (\ref{eq2: tau})
in terms of the microscopic collision-quantities:
\begin{eqnarray} \label{eq2: tau2}
\tau (E)^{-1}  = \frac{\Delta E}{E} \frac{1}{\Delta t},
\end{eqnarray}
where $\Delta E$ is the difference between the initial and final
energies of the proton before and after each collision, and
$\Delta t$ is the characteristic time between collisions.
Introducing the inelasticity through its definition $K = \Delta E
/ E$, and expressing the characteristic time in terms of the
scattering cross section and density $\rho$ of the target photons,
we can write then:
\begin{eqnarray}
\tau (E)^{-1} = K \, \sigma \rho \, v_{\mathrm{rel}} ,
\end{eqnarray}
where $v_{\mathrm{rel}}$ is the relative velocity between the
incident proton and the background. The above relation can be
driven to a more accurate version if we consider that both
$v_{\mathrm{rel}}$ and $\sigma$ are functions of the energy and
direction of propagation of the CMBR photons relative to the
incident proton. Considering these elements we are able to write
\begin{eqnarray} \label{eq2: tau3}
d \tau (E)^{-1} = K \, \sigma \, v_{\mathrm{rel}} \eta(\omega) d
\omega \, d \Omega / 4 \pi,
\end{eqnarray}
where $\eta (\omega) d \omega$ is the CMBR density of photons with
energies in the range $[\omega$, $\omega + d \omega]$, and $d
\Omega / 4\pi = \sin \theta \, d \theta d \phi /4\pi$ is the
section of solid angle. With the above quantities, it is simple to
rewrite $v_{\mathrm{rel}}$ through
\begin{eqnarray} \label{eq2: vrel}
v_{\mathrm{rel}} \, d \Omega / 4 \pi =  \frac{\epsilon \, d
\epsilon \, d\phi}{4 \pi \gamma^{2} \omega^{2}},
\end{eqnarray}
with $\epsilon \in [0,2 \gamma \omega]$ and $\phi \in [0,2 \pi]$.
Substituting equation (\ref{eq2: vrel}) in (\ref{eq2: tau3}) and
using the fact that the CMBR density corresponds to a Planck
distribution $\eta(\omega) d \omega = \omega^2 d \omega / \pi^2
(e^{\omega /kT}-1)$, it is finally possible to show that the mean
life $\tau (E)$ can be written in the form
\begin{eqnarray} \label{eq2: tau5}
\tau (E)^{-1}= -\frac{kT}{2 \pi^{2} \gamma^{2}} \!
\int_{\epsilon_{\mathrm{th}}}^{\infty} \!\!\!\! d \epsilon \,
\sigma (\epsilon) K(\epsilon)  \epsilon  \ln [1-e^{-\epsilon/ 2
\gamma k T}].
\end{eqnarray}

\subsection{Energy Loss and Spectrum}

The energy loss suffered by a very energetic proton during its
journey, from a distant source to our detectors, is not only
produced by the collisions that it has with CMBR at a particular
epoch. There will also be a decrease in its energy due to the
redshift attenuation produced by the expansion of the Universe. At
the same time, such expansion will affect the collision rate
through the attenuation of the photon gas density, which can be
understood as a cooling of the CMBR through the relation $T =
(1+z) T_{0}$, where $z$ is the redshift and $T_{0}$ is the
temperature of the background at the present time. To calculate
the spectrum we need to consider the rate of energy loss during
any epoch $z$ of the Universe.

For the present discussion, we shall assume that the Universe is
well described by a matter dominated Friedman Robertson Walker
(FRW) space-time, and that the ratio of density $\Omega_{0} =
\rho/\rho_{c}$ (where $\rho$ is the energy density of the present
Universe and $\rho_{c}$ is the critical energy density for the
Universe to be flat) is such that $\Omega_{0}=1$. The above
assumptions give rise to the following relation between the
temporal coordinate $t$ (proper time in the co-moving system) and
the redshift $z$:
\begin{eqnarray} \label{eq2: t-z flat universe}
dt = - \frac{dz}{H_{0} (1+z)^{5/2}},
\end{eqnarray}
where $H_{0}$ is the Hubble constant at present time. Since the
momentum of a free particle in a FRW space behaves as $p \propto
(1 + z)$, we will have, with the additional consideration $p \gg
m$ (where $m$ is the particle mass), that the energy loss due to
redshift is
\begin{eqnarray} \label{eq2: redshift contribution}
\left( -\frac{1}{E} \frac{dE}{dt} \right)_{\mathrm{cr}} = H_{0}
(1+z)^{3/2}.
\end{eqnarray}

On the other hand, the energy loss due to collisions with the CMBR
will evolve as the background temperature changes (recall $T =
(1+z) T_{0}$). This evolution can be parameterized through $z$ and
is given by
\begin{eqnarray} \label{eq2: Evol of Gamma}
\left( -\frac{1}{E} \frac{dE}{dt} \right)_{\mathrm{col}} =
(1+z)^{3} \tau ([1+z]E)^{-1}.
\end{eqnarray}

The total energy loss can be expressed as the addition of the
former contributions (using $z$ instead of $t$)
\begin{eqnarray} \label{eq2: diff E-z}
\frac{1}{E} \frac{dE}{dz} = (1+z)^{-1} + H_{0}^{-1} (1+z)^{1/2}
\tau ([1+z]E)^{-1}.
\end{eqnarray}
Equation (\ref{eq2: diff E-z}) can be numerically integrated to
give the energy $E_{g}(E,z)$ of a proton generated by the source
in a $z$ epoch and that will be detected with a energy $E$ here on
Earth. Let us designate this solution by the formal expression
\begin{eqnarray} \label{eq2: sol formal Eg}
E_{g}(E,z)=\lambda(E,z) E.
\end{eqnarray}

It is also possible to manipulate equation (\ref{eq2: diff E-z})
to obtain an expression for the dilatation of the energy interval
$dE_{g}/dE$. To accomplish this it is necessary to integrate
(\ref{eq2: diff E-z}) with respect to $z$ and then differentiate
it with respect to $E$ to obtain an integral equation for $d E_{g}
/ dE$. The solution of such an equation is found to be
\begin{eqnarray} \label{eq: dE-dE}
\frac{dE_{g}(z_{g})}{dE} = (1+z_{g}) \qquad \qquad \qquad \qquad \nonumber\\
\times \exp \left[ \int^{z_{g}}_{0} \frac{dz}{H_{0}} (1+z)^{1/2}
\frac{db(E')}{dE'} \right].
\end{eqnarray}
where $E'=(1+z)\lambda (E,z) E$.

The total flux $dj(E)$ of emitted particles from a volume element
$dV=R^{3}(z)r^{2} dr d \Omega$, in the epoch $z$ and coordinate
$r$, measured from  Earth at present with energy $E$ is
\begin{eqnarray} \label{eq2: dj}
dj(E)dE = \frac{F(E_{0},z) dE_{0} n(z) dV}{(1+z) 4 \pi R_{0}^{2}
r^{2}},
\end{eqnarray}
where $j(E)$ is the particle flux per energy, $F(E_{0},z) dE_{0}$
the emitted particle flux within the range $(E_{0}, E_{0} +
dE_{0})$, and $n(z)$ the density of sources in $z$. As previously
mentioned, it is convenient to study the emission flux with a
power law spectrum of the type $F(E) \propto E^{-\gamma_{g}}$. It
can be shown that with such assumption, the relation between the
emission flux and the total luminosity $L_{p}$ of the source is
$F(E) = (\gamma_{g} - 2) L_{p} E^{-\gamma_{g}}$. To describe the
evolution of the sources we shall also use a power law behavior.
This will be done through the relation
\begin{eqnarray}
L_{p}(z)n(z) &=& (1+z)^{(3+m)}L_{p}(0)n(0), \nonumber\\
&=& (1+z)^{(3+m)} \mathcal{L}_{0},
\end{eqnarray}
in such a way that $m=0$ corresponds to the case in which sources
do not evolve. If we consider that $R_{0} = (1+z) R(z)$ and
$R(z)dr=dt$ for flat spaces (and $v \simeq 1$ for very energetic
particles), using (\ref{eq2: t-z flat universe}) to express all in
terms of $z$ and, finally, integrating (\ref{eq2: dj}) from $z=0$
to some $z=z_{max}$ for which sources are not relevant for the
phenomena, it is possible to obtain
\begin{eqnarray} \label{eq2: j}
j(E) = (\gamma_{g}-2) \frac{1}{4 \pi}
\frac{\mathcal{L}_{0}}{H_{0}} E^{-\gamma_{g}} \qquad \qquad \nonumber\\
 \times  \int^{z_{\mathrm{max}}}_{0} \!\!\!\!\!\! dz_{g}
(1+z_{g})^{m-5/2} \lambda^{-\gamma_{g}}(E,z_{g})
\frac{dE_{g}(z_{g})}{dE}.
\end{eqnarray}
The above expression constitutes the spectrum of UHECR. It remains
to fix (observationally) the volumetric luminosity
$\mathcal{L}_{0}$ and the $\gamma_{g}$ and $m$ indices.

\subsection{Ultra High Energy Cosmic Rays Spectrum}

To accomplish the computation of the theoretical spectrum we need
information about the dynamical processes taking place in the
propagation of protons along the CMBR. As we already emphasized,
the most important reactions taking place in the description of a
proton's propagation are, the pair creation $ p + \gamma
\rightarrow p + e^{-} + e^{+}$, and the photo-pion production $p +
\gamma \rightarrow p + \pi$. This last reaction is mediated by
several channels. The main channels are
\begin{eqnarray}
p + \gamma & \rightarrow & N + \pi      \label{eq2: N+pi}\\
& \rightarrow & \Delta + \pi            \label{eq2: Delta+pi}\\
& \rightarrow & R                        \label{eq2: Res}\\
& \rightarrow & N + \rho(770)               \label{eq2: N+rho}\\
& \rightarrow & N + \omega(782).          \label{eq2: N+omega}
\end{eqnarray}
The total cross sections and inelasticities of these processes are
well known and can be used in (\ref{eq2: tau5}) to compute the
main time life of protons as a function of their energy. Then,
with the help of expressions (\ref{eq: dE-dE}) and (\ref{eq2: j}),
we can finally find the predicted spectrum for UHECR.

\begin{figure}[ht] 
\begin{center}
\includegraphics[width=0.48\textwidth]{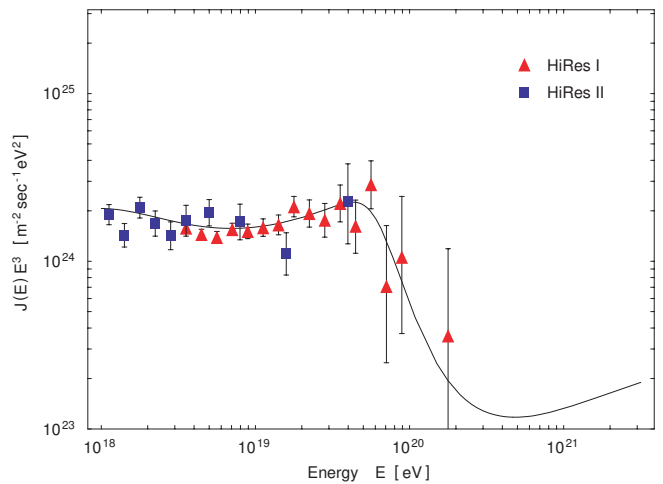}
\caption[Cosmic ray spectrum and HiRes]{UHECR spectrum and HiRes
observations. The figure shows the UHECR spectrum $J(E)$
multiplied by $E^{3}$, for uniform distributed sources, without
evolution ($m=0$), generation index $\gamma_{g} = 2.7$, and with a
maximum generation energy $E_{\mathrm{max}} = \infty$. Also shown
are the HiRes observed events.} \label{F1}
\end{center}
\end{figure}
\begin{figure}[ht] 
\begin{center}
\includegraphics[width=0.48\textwidth]{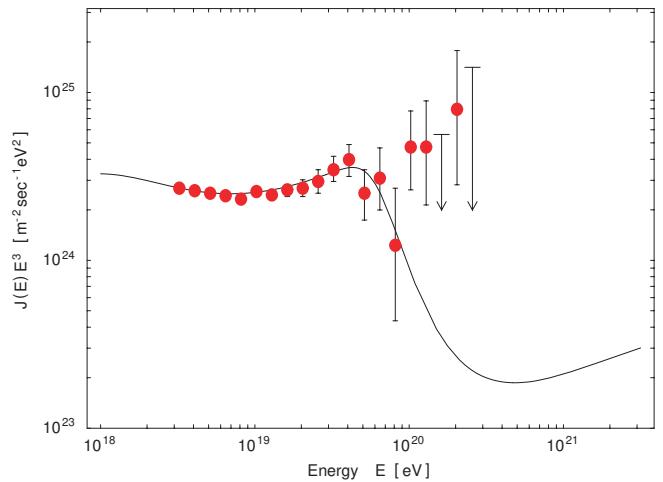}
\caption[Cosmic ray spectrum and AGASA]{UHECR spectrum and AGASA
observations. The figure shows the UHECR spectrum $J(E)$
multiplied by $E^{3}$, for uniform distributed sources, without
evolution, and with a maximum generation energy $E_{\mathrm{max}}
= \infty$. Also shown are the AGASA observed events. The best fit
for the low energy sector ($E < 4 \times 10^{19}$ eV) corresponds
to $\gamma_{g} = 2.7$.} \label{F2}
\end{center}
\end{figure}
 Fig.\ref{F1} shows the obtained spectrum $J(E)$ of UHECR and
the HiRes observed data (two detectors, HiRes-I and HiRes-II). In
order to emphasize the appearance of the GZK-cutoff in the
spectrum, we have selected the idealized case when the maximum
generation energy $E_{\mathrm{max}}$ for the emitted particles
from sources, is $E_{\mathrm{max}} = \infty$. To fit the HiRes
data, the generation index for the theoretical spectrum shown in
the figure is $\gamma_{g} = 2.7$, while the evolution index is
$m=0$. Additionally, the volumetric luminosity is $\mathcal{L}_{0}
= 2.96 \times 10^{51} \mathrm{ergs}/\mathrm{Mpc}^{3}\mathrm{yr}$.

Fig.\ref{F2} shows the obtained spectrum $J(E)$ of UHECR and the
AGASA observed data. Again, we have selected for the theoretical
spectrum $J(E)$ the idealized case $E_{\mathrm{max}} = \infty$. To
reconcile the data of the low energy region ($E < 4 \times
10^{19}$ eV), where the pair creation dominates the energy loss,
it is necessary to have a generation index $\gamma_{g} = 2.7$
(with the additional supposition that sources do not evolve) and a
volumetric luminosity $\mathcal{L}_{0} = 4.7 \times 10^{51}
\mathrm{ergs}/\mathrm{Mpc}^{3}\mathrm{yr}$. It can be seen that
for events with energies $E > 4 \times 10^{19}$ eV, where the
energy loss is dominated by the photo-pion production, the
predicted spectrum does not fit the data well. To have a
statistical sense of the discrepancy between observation and
theory, we can calculate the Poisson probability $P$ of an excess
in the five highest energy bins. This is $P = 1.1 \times 10^{-8}$.
Another statistical measure is provided by the Poisson $\chi^{2}$
given by \cite{PDG}
\begin{eqnarray} \label{eq2: poisson chi}
\chi^{2} = \sum_{i} [ 2 ( N_{i}^{\mathrm{th}} -
N_{i}^{\mathrm{obs}} ) + 2 N_{i}^{\mathrm{obs}} \ln (
N_{i}^{\mathrm{obs}} / N_{i}^{\mathrm{th}} ) ].
\end{eqnarray}
Computing this quantity for the eight highest energy bins, we
obtain $\chi^{2} = 29$. These quantities show how far the AGASA
measurements are from the theoretical prediction given by the
curve of Fig.\ref{F2}. Other more sophisticated models have also
been analyzed in detail \cite{Berezinsky}, nevertheless, it has
turned out that conventional physics does not have the capacity to
reproduce the observations from the AGASA collaboration group in a
satisfactory way.

 Whether HiRes or AGASA data are pointing in the right
direction to describe the correct pattern present in the arrival
of UHECR is still an open issue (see, for example, reference
\cite{MBO} for a detailed comparison between both experimental
results). In the rest of the paper we shall focus our attention on
the possibility of an absence of the GZK cutoff as a consequence
of LQG effects. For this reason, we will later return to the AGASA
observations in order to contrast the results of the following
sections.


\section{Loop Quantum Gravity} \label{LQG}

Loop quantum gravity is a canonical approach to the problem of
gravity quantization. It is based on the construction of a spin
network basis, labelled by graphs embedded in a three dimensional
insertion $\Sigma$ in space-time. A consequence of this approach
is that the quantum structure of space-time will be of a
polymer-like nature, highly manifested in phenomena involving the
Planck scale $l_{p}$.

The above, and very brief outline of loop quantum gravity allows
us to figure out how complicated a full treatment of a physical
phenomena could be when the quantum nature of gravity is
considered, even if the physical system is characterized by a flat
geometry. It is possible, however, to introduce a loop state which
approximates a flat 3-metric on $\Sigma$ at length scales greater
than the length scale $\mathcal{L} \gg l_{p}$. For pure gravity,
this state is referred to as the weave state $|W \rangle$, and the
length scale $\mathcal{L}$ as the weave scale. A flat weave $|W
\rangle$ will be characterized by $\mathcal{L}$ in such a way that
for distances $d \ll \mathcal{L}$ the quantum loop structure of
space is manifest, while for distances $d \geqslant \mathcal{L}$
the continuous flat geometry is regained. With this approach, for
instance, the metric operator $\hat q_{a b}$ satisfies
\begin{eqnarray} \label{eq3: weave-g}
\langle W| \hat q_{a b} |W \rangle = \delta_{ab} + \mathcal{O}
(l_{p} / \mathcal{L}).
\end{eqnarray}

A generalization of the former idea, to include matter fields, is
also possible. In this case, the loop state represents a matter
field $\psi$ coupled to gravity. Such a state is denoted by $|W,
\psi \rangle$ and, again, is simply referred to as the weave. As
before, it will be characterized by the weave scale $\mathcal{L}$
and the hamiltonian operators $\hat H_{\psi}$ are expected to
fulfill a relation analog to (\ref{eq3: weave-g}), that is, we
shall be able to define an effective hamiltonian $H_{\psi}$ such
that
\begin{eqnarray}
H_{\psi}= \langle W,\psi| \hat H_{\psi} |W,\psi \rangle.
\end{eqnarray}
An approach to this task has been performed by J. Alfaro {\it et
al.} \cite{Neutrinos, Photons, AMU} for 1/2-spin fermions and the
electromagnetic field. In this approach the effects of the loop
structure of space at the Planck level are treated semiclassically
through a coarse-grained approximation \cite{Gambini Pullin}. This
method leads to the natural appearance of LIV's in the equations
of motion derived from the effective hamiltonian. The key feature
here is that the effective hamiltonian is constructed from
expectation values of dynamical quantities from both the matter
fields and the gravitational field. In this way, when a flat weave
is considered, the expectation values of the gravitational part
will appear in the equations of motion for the matter fields in
the form of coefficients with dependence in both scales,
$\mathcal{L}$ and $l_{p}$. When a flat geometry is considered, the
expectation values can be interpreted as vacuum expectation values
for the considered matter fields.

A significant discussion is whether the Lorentz symmetry is
present in the full LQG theory (as in its classical counterpart)
or not \cite{Thiemann1}. For the present work, we shall assume
that Lorentz symmetry is indeed present in the full LQG theory.
This assumption, jointly with the consideration that the new
corrective coefficients are vacuum expectation values, leads us to
consider that the Lorentz symmetry is spontaneously broken in the
effective theory level.

In what follows we will briefly summarize the obtained equations
of motion for both, 1/2-spin fermions and photons, as well as the
obtained dispersion relations. 

\subsection{Fermions} \label{fermions}

The LQG effective equations of motion for a $1/2$-spin fermion
field, coupled to gravity, are \cite{Neutrinos}
\begin{eqnarray}
\left[i \frac{\partial}{\partial t} - i \hat A \vec \sigma \cdot
\nabla + \frac{\hat{B}}{2 \mathcal{L}} \right] \xi(x) \qquad \qquad \qquad \nonumber\\
 - m (C - i D \vec \sigma \cdot \nabla)
\chi(x) = 0, \label{eq3: neumod1}\\
\left[i \frac{\partial}{\partial t} + i \hat A \vec \sigma \cdot
\nabla - \frac{\hat{B}}{2 \mathcal{L}} \right] \chi(x) \qquad \qquad \qquad \nonumber\\
 - m (C - i D \vec \sigma \cdot \nabla)
\xi(x)=0, \label{eq3: neumod2}
\end{eqnarray}
where $\xi (x)$ and $\chi (x)$ are the spinor components for the
Dirac field $\Psi(x) = (\xi (x) , \chi (x))$ and the hermitian
operators, $\hat A$ and $\hat C$, are given by the following
expressions
\begin{eqnarray} \hat A &=& 1 + \kappa_{1}
\frac{l_{p}}{\mathcal{L}} + \kappa_{2} \left(
\frac{l_{p}}{\mathcal{L}} \right)^{2} +
\frac{\kappa_{3}}{2} l_{p}^{2} \nabla^{2}, \\
\hat B &=& \kappa_{5} \frac{l_{p}}{\mathcal{L}} + \kappa_{6}
\left( \frac{l_{p}}{\mathcal{L}} \right)^{2} +
\frac{\kappa_{7}}{2} l_{p}^{2} \nabla^{2},
\end{eqnarray}
and the constants $C$ and $D$ are given by
\begin{eqnarray}
C &=& 1 + \kappa_{8} \frac{l_{p}}{\mathcal{L}}, \\
D &=& \frac{\kappa_{9}}{2 \hbar} l_{p}.
\end{eqnarray}
In the above expressions the $\kappa_{i}$ quantities are unknown
coefficients of order 1 which need to be determined. In the case
that $\Psi(x)$ is a Majorana field, the $\xi (x)$ and $\chi (x)$
spinors fulfill the reality condition:
\begin{eqnarray} \label{eq3: real condition}
\xi(x) = -i \sigma^{2} \, \chi^{*}(x) \quad \mathrm{and} \quad
\chi(x) = i \sigma^{2} \, \xi^{*}(x).
\end{eqnarray}
With the help of this condition, equations (\ref{eq3: neumod1})
and (\ref{eq3: neumod2}) can be simplified to
\begin{eqnarray} \label{eq3: neumod3}
\left[\frac{\partial}{\partial t} - \hat A \vec \sigma \cdot
\nabla - \frac{i \hat{B}}{2 \mathcal{L}} \right] \xi(x)  \qquad
\qquad \qquad
\nonumber\\
 - m (C - i  D \vec \sigma \cdot \nabla) \sigma^{2}
\xi^{*}(x)=0.
\end{eqnarray}
Equations (\ref{eq3: neumod1}) and (\ref{eq3: neumod2}) are
invariant under charge conjugation C and time inversion T, but not
under parity conjugation P. As a consequence, the fermion equation
of motion violates the CPT symmetry through P. The terms that
produce the P violation are those related with $\hat B$ and $D$.

Some comments need to be made at this stage. Of specially
importance to the development of the above effective equations of
motion is that they are only valid in a homogenous and isotropic
system. From the point of view of a spontaneous symmetry breakup
such a system is unique and, therefore, a privileged reference
frame). It is possible then to put the equations of motion (and
therefore the dispersion relations) in a covariant form through
the introduction of a four-velocity vector explicitly denoting the
existence of a preferred system. From the cosmological point of
view, such a privileged system does exist, and corresponds to the
CMBR co-moving reference system. For that reason, we shall assume
that the preferred system denoted by the presence of LIV's is the
same CMBR co-moving reference frame, and will use it as the
laboratory system.

The dispersion relation for fermions can be easily obtained
through the development of the Klein-Gordon-like equation. The
obtained dispersion relation is
\begin{eqnarray} \label{eq3: E fermion}
E_{\pm}^{2} = (Ap \pm \frac{B}{2 \mathcal{L}})^{2} + m^{2} (C \pm
D p)^{2}
\end{eqnarray}
where the $\pm$ signs correspond to the helicity state of the
described particle (note that these signs are produced by the
parity violation coefficients), and where now we have
\begin{eqnarray} \label{eq3: Def A,C,alpha,beta}
A &=& 1+ \kappa_{1} \frac{l_{p}}{\mathcal{L}}+\kappa_{2} \left(
\frac{l_{p}}{\mathcal{L}}  \right)^{2}
+\frac{\kappa_{3}}{2} l_{p}^{2} p^{2} , \nonumber \\
B &=& \kappa_{5} \frac{l_{p}}{\mathcal{L}}+\kappa_{6} \left(
\frac{l_{p}}{\mathcal{L}}  \right)^{2}
+\frac{\kappa_{7}}{2} l_{p}^{2} p^{2}, \nonumber \\
C &=& 1+ \kappa_{8} \frac{l_{p}}{\mathcal{L}}, \nonumber \\
D &=& \frac{\kappa_{9}}{2} l_{p}.
\end{eqnarray}
For our purposes, it will be sufficient to consider the lower
contributions in both scales, $l_{p}$ and $\mathcal{L}$
\cite{Alfaro & Palma}
\begin{eqnarray} \label{eq: E for fermions}
E^{2}_{\pm} = p^{2} + 2 \alpha p^{2} + \eta p^{4} \pm 2 \lambda p
+ m^{2},
\end{eqnarray}
where we have defined the new set of corrections $\alpha$, $\eta$
and $\lambda$ depending on the scales $\mathcal{L}$ and $l_{p}$ in
the following way:
\begin{eqnarray}
\alpha &=& \kappa_{\alpha} (l_{p} / \mathcal{L})^{2} \\
\eta &=& \kappa_{\eta} \, l_{p}^{2} \\
\lambda &=& \kappa_{\lambda} \, l_{p} / 2 \mathcal{L}^{2},
\end{eqnarray}
being $\kappa_{\alpha}$, $\kappa_{\eta}$ and $\kappa_{\lambda}$
adimensional parameters of order 1.

\subsection{Photons}

For the electromagnetic sector of the theory we have the following
set of effective equations
\begin{eqnarray}
A (\nabla \times \vec B) - \frac{1}{c} \frac{\partial \vec
E}{\partial t} + 2 l_{p}^{2} \theta_{3} \nabla^{2} (\nabla \times
\vec B) \nonumber\\
- 2 \theta_{8} l_{p} \nabla^{2} \vec B + 4 \theta_{4}
\mathcal{L}^{2} \left( \frac{\mathcal{L}}{l_{p}} \right)^{2
\Upsilon} l_{p}^{2} \nabla \times (\vec B^{2} \vec B) = 0, \label{eq3: electromag1}\\
A(\nabla \times \vec E) + \frac{1}{c} \frac{\partial \vec
B}{\partial t} + 2 l_{p}^{2} \theta_{3} \nabla^{2} (\nabla \times
\vec E) \nonumber\\
- 2 \theta_{8} l_{p} \nabla^{2} \vec E = 0, \label{eq3:
electromag2}
\end{eqnarray}
where
\begin{eqnarray}
A = 1 + \theta_{7} \left( \frac{l_{p}}{\mathcal{L}} \right)^{2 + 2
\Upsilon}.
\end{eqnarray}

To calculate a dispersion relation for photons we need to consider
only the linear part of equations (\ref{eq3: electromag1}) and
(\ref{eq3: electromag2}), and try solutions of the type $\vec E =
\vec E_{0} e^{i(\vec k \cdot \vec x - \omega t)}$ and $\vec B =
\vec B_{0} e^{i(\vec k \cdot \vec x - \omega t)}$, for the
electric and magnetic fields. In this way, the obtained dispersion
relation between the energy $\omega$ and the momentum $k$ of
photons is
\begin{eqnarray} \label{eq3: complet E for photons}
\omega_{\pm} = k [A_{\gamma} - \theta_{3}(l_{p} k)^{2} \pm
\theta_{8} l_{p} k],
\end{eqnarray}
where
\begin{eqnarray}
A_{\gamma} &=& 1 + \kappa_{\gamma} \left(
\frac{l_{p}}{\mathcal{L}} \right)^{2+2\Upsilon}.
\end{eqnarray}
In the previous expression the $\kappa_{\gamma}$ and $\theta_{i}$
coefficients are adimensional parameters of order 1. As before,
the $\pm$ signs refer to the helicity state of the described
photons. The $\Upsilon$ quantity is a free parameter that measures
a possible non-canonical scaling of the gravitational expectation
values in the semiclassical state (let us note that the presence
of $\Upsilon$ in the fermionic sector was not considered in
\cite{Neutrinos}). To be consistent with the dispersion relation
of fermions, we shall consider only possibilities $\Upsilon$ =
-1/2, 0, 1/2, 1, etc., in such a way that $A_{\gamma} \sim 1 +
\mathcal{O}[(l_{p}/\mathcal{L})^{n}]$, where $n=2+2\Upsilon$ is a
positive natural number. With this supposition, we can find a
tentative value for $\Upsilon$, through the bound of the lower
order correction $\delta A \sim \mathcal{O}[(l_{p} /
\mathcal{L})^{n}]$ (where $\delta A = A_{\gamma} - A_{a}$, being
$a$ another particle).

Considering the lower order contributions in both scales, $l_{p}$
and $\mathcal{L}$, we are able to simplify the photon dispersion
relation to
\begin{eqnarray} \label{eq3: E for photons}
\omega^{2}_{\pm} = k^{2} + 2 \alpha_{\gamma} k^{2} \pm 2
\theta_{\gamma} l_{p} k^{3},
\end{eqnarray}
where $\alpha_{\gamma}$ is defined by the relation $A_{\gamma} = 1
+ \alpha_{\gamma} = 1 + \kappa_{\gamma} \, (l_{p} /
\mathcal{L})^{2+2\Upsilon}$.

\subsection{Other Particles}

We have so far examined the dispersion relations coming from LQG
for both $1/2$-spin fermions and photons. A relevant issue for the
following developments is the establishment of a valid extension
of the former results for other particles. In particular, we are
interested in counting with dispersion relations for 3/2-spin
fermions and 0-spin massive bosons. A precise and rigorous
procedure would require a complete calculation of effective field
equations of motion coming from LQG for each particle flavor in
which we are interested. For present purposes we will assert that
the valid dispersion relation for more general fermions is simply
\begin{eqnarray} \label{eq3: E for fermions general}
E^{2}_{\pm} = p^{2} + 2 \alpha p^{2} + \eta p^{4} \pm 2 \lambda p
+ m^{2},
\end{eqnarray}
This assertion preserves the basic symmetries and assumptions that
have lead to the obtention of the equations of motion for 1/2-spin
fermions.

On the other hand, in the case of bosonic 0-spin particles, we
will assert that the valid dispersion relation consists of
\begin{eqnarray} \label{eq3: E for bosones spin 0}
E^{2} = p^{2} + 2 \alpha p^{2} + \eta p^{4} + m^{2}.
\end{eqnarray}
This assertion is based on the fact that the symmetries involved
in the construction of the effective hamiltonian for 0-spin bosons
would prevent the appearance of terms like $\lambda$ (which
depends on the helicity).

To conclude, let us mention that the dispersion relation
(\ref{eq3: E for fermions general}) will be used for the physical
description of electrons, protons, neutrons, and $\Delta$ and $N$
baryonic resonances. Meanwhile, the dispersion relation (\ref{eq3:
E for bosones spin 0}) will be used for mesons $\pi$, $\rho$ and
$\omega$.


\section{Threshold Conditions} \label{Trh}

A useful discussion around the effects that LIV's can have on the
propagation of UHECR can be raised through the study of the
threshold conditions for the reactions to take place \cite{Coleman
& Glashow}. To simplify our subsequent discussions, let us use the
following notation for the modified dispersion relations
\begin{eqnarray}
E^{2} = p^{2} + f(p) + m^{2},
\end{eqnarray}
where $f(p)$ is the deformation function of the momentum $p$.

A decay reaction is kinematically allowed when, for a given value
of the total momentum $\vec p_{0} = \sum_{\mathrm{initial}} \vec p
= \sum_{\mathrm{final}} \vec p$, one can find a total energy value
$E_{0}$ such that $E_{0} \geqslant E_{\mathrm{min}}$. Here
$E_{\mathrm{min}}$ is the minimum value attainable by the total
energy of the decaying products for a given total momentum $\vec
p_{0}$. To find $E_{\mathrm{min}}$, it is enough to take the
individual decay product momenta to be collinear with respect to
the total momentum $\vec p_{0}$ and with the same direction. To
see this, we can variate $E_{0}$ with the appropriate restrictions
\begin{eqnarray}
E_{0} = \sum_{i} E_{i} (p_{i}) + \xi_{j} (p_{0}^{\, j} - \sum_{i}
p_{i}^{\, j}),
\end{eqnarray}
where $\xi_{j}$ are Lagrange multipliers, the $i$ index specifies
the $i$th particle and the $j$ index the $j$th vectorial component
of the different quantities. Doing the variation, we obtain
\begin{eqnarray}
\frac{\partial E_{i}}{\partial p_{i}^{j}} \equiv v_{i}^{j} =
\xi_{j}.
\end{eqnarray}
That is to say, the velocities of all the final produced particles
must be equal to $\xi$. Since the dispersion relations that we are
treating are monotonously increasing in the range of momenta $p >
\lambda$, the momenta can be taken as being collinear and with the
same direction of the initial quantity $\vec p_{0}$.

In this work, we will focus on those cases in which two particles
(say $a$ and $b$) collide to subsequently decay in the
aforementioned final states. For the present discussion, particles
$a$ and $b$ have momenta $\vec p_{a}$ and $\vec p_{b}$
respectively, and the total momentum of the system is $\vec
p_{0}$. It is easy to see from the dispersion relations that we
are considering, that the total energy of the system will depend
only on $p_{a} = |\vec p_{a}|$ and $p_{b} = |\vec p_{b}|$.
Therefore, to obtain the threshold condition for the mentioned
kind of process, we must find the maximum possible total energy
$E_{\mathrm{max}}$ of the initial configuration, given the
knowledge of $p_{a}$ and $p_{b}$. To accomplish this, let us fix
$\vec p_{a}$ and variate the incoming direction of $\vec p_{b} =
\hat n p_{b}$ in
\begin{eqnarray} \label{eq: restricion 2 over E}
E_{0} & = & E_{a}(\vec p_{0} - p_{b} \hat n) + E_{b} (p_{b}) +
\chi (\hat n ^{2} - 1).
\end{eqnarray}
Varying (\ref{eq: restricion 2 over E}) with respect to $\hat n$
($\chi$ is a Lagrange multiplier), we find
\begin{eqnarray}
\hat n ^{i} = \frac{v_{a}^{i} p_{b}}{2 \chi}.
\end{eqnarray}
In this way we obtain two extremal situations $\chi = \pm v_{a}
p_{b}/2$, or simply
\begin{eqnarray}
\hat n ^{i} = \pm \frac{v_{a}^{i}}{v_{a}}.
\end{eqnarray}
A simple inspection shows that for the dispersion relations that
we are considering, the maximum energy is given by $\hat n ^{i} =
- v_{a}^{i}/v_{a}$, or in other words, when a frontal collision
takes place.

Summarizing the threshold condition for a two particle ($a$ and
$b$) collision and subsequent decay, can be expressed through the
following requirements:
\begin{eqnarray}
E_{a} + E_{b} \geqslant \sum_{\mathrm{final}} E_{f},
\end{eqnarray}
with all final particles having the same velocity ($v_{i} = v_{j}$
for any to final particles $i$ and $j$), and
\begin{eqnarray}
p_{a} - p_{b}  = \sum_{\mathrm{final}} p_{f},
\end{eqnarray}
where the sign of the momenta $\sum_{\mathrm{final}} p_{f}$ is
given by the direction of the highest momentum magnitude of the
initial particles.

Our interest in the next subsections is the study of the reactions
involved in high energy cosmic ray phenomena through the threshold
conditions. To accomplish this goal through simple expressions
that are easy-to-manipulate, we shall further use, for the equal
velocities condition, the simplification
\begin{eqnarray}
E_{b}  m_{a} = E_{a}  m_{b},
\end{eqnarray}
valid for the study of parameters coming from the region $f(p) \ll
m^{2}$. This simplification will allow the achievement of bounds
over the order of magnitude of the different parameters involved
in the modified dispersion relations, which are precisely our main
concern.

In the following subsections we will study the kinematical effects
of LIV's through the threshold conditions for the reactions
involved in the propagation of UHECR. Since, in this phenomena,
photons are present in the form of low energy particles (the soft
photons of the CMBR), the LQG corrections in the electromagnetic
sector of the theory can be ignored. LQG corrections to the
electromagnetic sector, however, have already been studied for
other high energy reactions such as the Mkn 501 $\gamma$-rays
\cite{Alfaro & Palma}.

\subsection{Photo-Pion Production $\gamma
+ p \rightarrow p + \pi$}

Let us begin with the photo-pion production $\gamma + p
\rightarrow p + \pi$. Considering the corrections provided in the
dispersion relations (\ref{eq: E for fermions}) and (\ref{eq3: E
for bosones spin 0}) for fermions and bosons, we note that, for
the photo-pion production to proceed, the following condition must
be satisfied
\begin{eqnarray} \label{eq4: umbral pi}
2 \, \delta \alpha \, E_{\pi}^{2} + \left( \delta \eta + 3
\eta_{p}
\frac{m_{p} ( m_{p}+m_{\pi})}{m_{\pi}^{2}} \right) E_{\pi}^{4} \nonumber\\
 + 2 E_{\pi} (|\lambda_{p}| \pm \lambda_{p}) + 4E_{\pi} \omega \geqslant
\frac{m_{\pi}^{2}(2 m_{p}+m_{\pi})}{m_{p}+m_{\pi}},
\end{eqnarray}
where $E_{\pi}$ is the energy of the emergent pion, $\delta \alpha
= \alpha_{p} - \alpha_{\pi}$ and $\delta \eta = \eta_{p} -
\eta_{\pi}$. In expression (\ref{eq4: umbral pi}), the $\pm$ signs
refer to the helicity of the incident proton. Since there will
necessarily be a proton helicity that can minimize the term
associated with $\lambda_{p} \,$ and, therefore, minimize the
energy configuration for the threshold condition, we must insert,
in (\ref{eq4: umbral pi}), the following equality
\begin{eqnarray}
2 E_{\pi} (|\lambda_{p}| \pm \lambda_{p}) = 0.
\end{eqnarray}
In addition, we are assuming that the difference between $\kappa$
parameters from different particles are of order 1 ($\delta \kappa
\sim 1$). Therefore, if not null, we can take $\eta_{p}$ to
dominate over $\delta \eta$ in (\ref{eq4: umbral pi}). With these
considerations in mind, we are left with
\begin{eqnarray} \label{eq: pion}
2 \, \delta \alpha \, E_{\pi}^{2} + 168 \eta_{p} E_{\pi}^{4} +
4E_{\pi} \omega \geqslant \frac{m_{\pi}^{2}(2
m_{p}+m_{\pi})}{m_{p}+m_{\pi}}.
\end{eqnarray}
Note that in the absence of LQG corrections, the threshold
condition is simply
\begin{eqnarray}
4E_{\pi} \omega \geqslant \frac{m_{\pi}^{2}(2
m_{p}+m_{\pi})}{m_{p}+m_{\pi}}.
\end{eqnarray}

\subsection{Resonant Production $\gamma + p \rightarrow \Delta$}

The main channel involved in the photo-pion production is the
resonant production of the $\Delta (1232)$. It can be shown that
the threshold condition for the resonant $\Delta (1232)$ decay
reaction to occur, is
\begin{eqnarray} \label{Delta}
2 \, \delta \alpha \, E_{p}^{2} + \delta \eta \, E_{p}^{4}
 + 2 \left[  (\pm)_{p} \, \lambda_{p} +
|\lambda_{\Delta}| \, \right] E_{p} \nonumber\\ + 4 \omega E_{p}
\geqslant m_{\Delta}^{2} - m_{p}^{2} \, ,
\end{eqnarray}
where $E_{p}$ is the incident proton energy, $\delta \alpha =
\alpha_{p} - \alpha_{\Delta}$ and $\delta \eta = \eta_{p} -
\eta_{\Delta}$. Additionally, $(\pm)_{p}$ refers to the incident
proton helicity. In the absence of LQG corrections, the
conventional threshold condition is naturally reobtained:
\begin{eqnarray} \label{eq4: GZK class}
E_{p} \geqslant \frac{m_{\Delta}^{2} - m_{p}^{2}}{4 \omega}.
\end{eqnarray}

\subsection{Pair Creation $\gamma + p \rightarrow p + e^{+} + e^{-}$}

Pair creation, $\gamma + p \rightarrow p + e^{+} + e^{-}$, is
greatly abundant in the sector previous to the GZK limit. When the
dispersion relations for fermions are considered for both protons
and electrons, it is possible to find
\begin{eqnarray} \label{pair creation}
\delta \alpha \frac{m_{e}}{m_{p} + 2 m_{e}} E^{2} \qquad \qquad \qquad
\qquad \qquad \qquad \qquad \nonumber\\
+ 2 \left( \delta \eta + \frac{3}{4} \eta_{p} \frac{m_{p} ( m_{p} + 2
m_{e})}{m_{e}^{2}} \right)
\left( \frac{m_{e}}{m_{p} + 2 m_{e}} \right)^{3} E^{4}   \nonumber\\
+ E \omega + |\lambda_{e}| E + \frac{1}{2} (|\lambda_{p}| \pm
\lambda_{p}) E  \geqslant  m_{e} (m_{p} + m_{e}),
\end{eqnarray}
with $\delta \alpha = \alpha_{p} - \alpha_{e}$ and $\delta \eta =
\eta_{p} - \eta_{e}$.

As in the case of photo-pion production, there will always be an
incident proton helicity which can minimize the inequality
(\ref{pair creation}). Therefore, to study the production of the
electron-positron pair under its threshold condition, we shall set
$|\lambda_{p}| \pm \lambda_{p} = 0$. On the other hand, since our
intention is to estimate an order of magnitude for the value of
the diverse parameters present in the theory, let us ignore the
$\delta \eta$ term, since the presence of $\eta_{p}$ is of greater
relevance (recall that we are considering that $\mathcal{O}(\eta)
= \mathcal{O}(\delta \eta)$). With these considerations, we obtain
\begin{eqnarray} \label{eq: pair}
\delta \alpha \, \frac{m_{e}}{m_{p}} E^{2} + \frac{3}{2} \eta
\frac{m_{e}}{m_{p}} E^{4} + |\lambda_{e}| E \nonumber\\ + E \omega
\geqslant m_{e} (m_{p} + m_{e}),
\end{eqnarray}
where we have also used $m_{p} + m_{e} \simeq m_{p}$, to simplify
the above expression.

Finally, if no corrections are present at all, the threshold
condition would be reduced to the conventional one,
\begin{eqnarray}
E \omega  \geqslant  m_{e} (m_{p} + m_{e}).
\end{eqnarray}

\subsection{Bounds}

In order to study the threshold conditions (\ref{eq: pion}),
(\ref{Delta}) and (\ref{eq: pair}), in the context of the GZK
anomaly, we must establish some criteria.

Firstly, as we have seen in section \ref{UHECR}, the
conventionally obtained theoretical spectrum provides a very good
description of the phenomena up to an energy $\sim 4 \times
10^{19}$ eV. The main reaction taking place in this well described
region is the pair creation $\gamma + p \rightarrow p + e^{+} +
e^{-}$ and, therefore, no modifications are present for this
reaction up to $\sim 4 \times 10^{19}$ eV. As a consequence, and
since threshold conditions offer a measure of how modified
kinematics is, we will require that the threshold condition
(\ref{eq: pair}) for pair creation not be substantially altered by
the new corrective terms.

Secondly, we have the GZK anomaly itself, which we are committed
to explain. Since for energies greater than $\sim 8 \times
10^{19}$ eV the conventional theoretical spectrum does not fit the
experimental data well, we shall require that LQG corrections be
able to offer a violation of the GZK-cutoff. The dominant reaction
in the violated $E > 8 \times 10^{19}$ region is the photo-pion
production and, therefore, we shall require further that the new
corrective terms present in the kinematical calculations be able
to shift the threshold significantly to preclude the reaction.

As a last possibility, we shall also examine the bounds arising
for the case in which no GZK anomaly (and therefore no violations
to the threshold and kinematics) really exists. Since the HiRes
data have reached the $\sim 1.8 \times 10^{20}$ eV, we will
consider the scenario in which no violation at all is confirmed by
the data up to a reference energy $E_{\mathrm{ref}} = 2 \times
10^{20}$ eV.

In order to study the different corrections, given that we don't
have a detailed knowledge of the deviation parameters, we shall
take account of them independently. Naturally, there will always
exist the possibility of having an adequate combination of these
parameter values that could affect the threshold conditions
simultaneously. However, as shall soon be evident, each one of
these parameters will be significant at different energy ranges.

\subsubsection{$\alpha$ Correction}

We shall begin our analysis with correction $\alpha$ and the
consideration of the threshold condition for pair production. In
this case we have
\begin{eqnarray}
\delta \alpha \, \frac{m_{e}}{m_{p}} E^{2} + E \omega \geqslant
m_{e} (m_{p} + m_{e}),
\end{eqnarray}
with $\delta \alpha = \alpha_{p} - \alpha_{e}$. As is clear from
the above condition, the minimum soft-photon energy
$\omega_{\mathrm{min}}$ for the pair production to occur, is
\begin{eqnarray}
\omega_{\mathrm{min}} = \frac{m_{e}}{E} (m_{p} + m_{e}) - \delta
\alpha \, \frac{m_{e}}{m_{p}} E.
\end{eqnarray}
It follows therefore that the condition for a significant increase
or decrease in the threshold energy for pair production becomes
$|\delta \alpha| \geqslant m_{p} (m_{p} + m_{e}) / E^{2}$. In this
way, if we do not want kinematics to be modified up to a reference
energy $E_{\mathrm{ref}} = 3 \times 10^{19}$, we must impose the
following constraint
\begin{eqnarray} \label{eq: bound alpha}
|\alpha_{p} - \alpha_{e}| < \frac{(m_{p} + m_{e} ) m_{p} }{
E_{\mathrm{ref}}^{2}} = 9.8 \times 10^{-22}.
\end{eqnarray}
Similar treatments can be found for the analysis of other
astrophysical signals like the Mkn 501 $\gamma$-rays \cite{Stecker
& Glashow}, when the absence of anomalies is considered.

Let us now consider the threshold condition for the photo-pion
production. Taking only the $\alpha$ correction, we have
\begin{eqnarray} \label{eq: pion2}
2 \, \delta \alpha \, E_{\pi}^{2} + 4E_{\pi} \omega \geqslant
\frac{m_{\pi}^{2}(2 m_{p}+m_{\pi})}{m_{p}+m_{\pi}}.
\end{eqnarray}
It is possible to find that for the above condition to be violated
for all energies $E_{\pi}$ of the emerging pion, and therefore no
reaction to take place, the following inequality must hold
\begin{eqnarray} \label{eq: pi-p}
\alpha_{\pi} - \alpha_{p} > \frac{2 \omega^{2} (m_{p} +
m_{\pi})}{m_{\pi}^{2} (2 m_{p} + m_{\pi})} = 3.3 \times 10^{-24}
\left[ \omega / \omega_{0} \right] ^{2}.
\end{eqnarray}
where $\omega_{0} = K T = 2.35 \times 10^{-4}$ eV is the thermal
CMBR energy. If we repeat these steps for the $\Delta(1232)$
resonant decay, we obtain the following condition
\begin{eqnarray}
\alpha_{\Delta} - \alpha_{p} >  \frac{2 \omega^{2}}{m_{\Delta}^{2}
- m_{p}^{2}} = 1.7 \times 10^{-25} \left[ \omega / \omega_{0}
\right] ^{2}.
\end{eqnarray}

To estimate a range for the weave scale $\mathcal{L}$, let us use
as a reference energy $\omega_{\mathrm{ref}} =
\omega_{\mathrm{min}}$, where $\omega_{\mathrm{min}}$ is the
minimum energy for the reaction to take place, in inequality
(\ref{eq: pion2}), when the condition for a significant increase
in the threshold condition is taken into account (for a primordial
proton reference energy $E_{\mathrm{ref}} = 2 \times 10^{20}$,
this is $\omega_{\mathrm{min}} \sim 2.9 \times \omega_{0}$), and
join the results deduced from the mentioned requirements. Assuming
that the $\kappa_{\alpha}$ parameters are of order 1, as well as
the difference between them for different particles, we can
estimate
---for the weave scale $\mathcal{L}$--- the preferred range
\begin{eqnarray} \label{eq: restriction}
2.6 \times 10^{-18} \,\,\, \mathrm{eV}^{-1} \lesssim \mathcal{L}
\lesssim 1.6 \times 10^{-17} \,\,\, \mathrm{eV}^{-1},
\end{eqnarray}
where the lefthand and righthand sides come from bounds (\ref{eq:
bound alpha}) and (\ref{eq: pi-p}) respectively (since the
$\Delta$(1232) is just one channel of the photo-pion production,
we shall not consider it to set any bound).

If no GZK anomaly is confirmed in future experimental
observations, then we should state a stronger bound for the
difference $\alpha_{\pi} - \alpha_{p}$. Using the same assumptions
to set the restriction (\ref{eq: bound alpha}) when the primordial
proton reference energy is  $E_{\mathrm{ref}} = 2 \times 10^{20}$
eV, it is possible to find
\begin{eqnarray}
|\alpha_{\pi} - \alpha_{p}| < 2.3 \times 10^{-23}.
\end{eqnarray}
In terms of the length scale $\mathcal{L}$, this last bound may be
read as
\begin{eqnarray}
\mathcal{L} \gtrsim 1.7 \times 10^{-17}  \,\, \mathrm{eV}^{-1},
\end{eqnarray}
which is a stronger bound over $\mathcal{L}$ than (\ref{eq: bound
alpha}), offered by pair creation.

\subsubsection{$\eta$ Correction}

Let us now turn our attention to the $\eta$ parameter. The
threshold condition for the pair production, when only the $\eta$
parameter is considered, is
\begin{eqnarray}
\frac{3}{2} \eta \frac{m_{e}}{m_{p}} E^{4}
 + E \omega \geqslant m_{e} (m_{p} + m_{e}).
\end{eqnarray}
Repeating the same analysis we did for the $\alpha$ parameter, it
is possible to find the following constraint
\begin{eqnarray} \label{eq4: cota eta-e}
|\eta| < \frac{2}{3} \frac{m_{p}}{E_{\mathrm{ref}}^{4}} (m_{p} +
m_{e}) = 1.6 \times 10^{-60} \,\, \mathrm{eV}^{-2}.
\end{eqnarray}
Recalling that $\eta = \kappa_{\eta} l_{p}^{2}$, result (\ref{eq4:
cota eta-e}) can be reexpressed in the form
\begin{eqnarray} \label{eq4: cota k2-e}
|\kappa_{\eta}| \lesssim 2.4 \times 10^{-4},
\end{eqnarray}
which is, of course, a strong bound over a parameter of order 1.

Since the basis of the effective LQG methods which we have
developed rely on the fact that the coefficients $\kappa$ are of
order 1, we must conclude that a reaction of the type $\eta$
should be discarded, in opposition to the expectations of our
previous work \cite{Alfaro & Palma}, when only photo-pion
production was analyzed.

\subsubsection{$\lambda$ Correction}

Finally we have to consider the $\lambda$ correction. In our
previous work \cite{Alfaro & Palma}, having studied the photo-pion
production through the $\Delta(1232)$ channel decay, we emphasized
the possibility of a helicity dependent violation. For this effect
to take place, the following configuration must be satisfied
(using again $\omega_{\mathrm{ref}} = \omega_{\mathrm{min}}$)
\begin{eqnarray} \label{lambda Delta}
|\lambda_{p}| \geqslant  |\lambda_{\Delta}| + 1.3 \times 10^{-3}
\,\, \mathrm{eV}.
\end{eqnarray}
However, when pair production is analyzed (in the same way as with
$\alpha$ and $\eta$), the following condition emerges for
$\lambda_{e}$:
\begin{eqnarray}
|\lambda_{e}| < 1.6 \times 10^{-5} \,\, \mathrm{eV},
\end{eqnarray}
which is more than 1 order of magnitude stronger than the required
value for protons in (\ref{lambda Delta}). This weakens the
possibility of a limit violation through helicity dependent
effects.

Another stronger bound can be found in \cite{Carmona Cortes},
where a dispersion relation of the type
\begin{eqnarray}
E^{2}=p^{2} + \lambda p + m^{2}
\end{eqnarray}
is analyzed, and it is found that the case $\lambda \geqslant 1
\times 10^{-7}$ eV should be discarded because of the highly
sensitive measurements of the Lamb shift.

\subsection{The Cubic Correction}

A commonly studied correction which has appeared in several recent
works \cite{Amelino, Ellis, Liberati1}, and which deserves our
attention, is the case of a cubic correction of the form
\begin{eqnarray} \label{eq4: lp-E3}
E^{2} = p^{2} + m^{2} + \xi \, p^{3},
\end{eqnarray}
(where $\xi$ is an arbitrary scale). It is interesting to note
that strong bounds can be placed over deformation $f(p) =
 \xi \, p^{3}$. We will assume in this section, that
$\xi$ is an universal parameter (an assumption followed by most of
the works in this field).

>From dispersion relation (\ref{eq4: lp-E3}), the threshold
condition for the photo-pion production is
\begin{eqnarray} \label{eq4: umbral xi - pi}
4 \, \omega E + 2 \, \xi \frac{m_{\pi} m_{p}}{ (m_{p} +
m_{\pi})^{2}} E^{3} \geqslant m_{\pi} (2 m_{p} + m_{\pi}).
\end{eqnarray}
Clearly, when $\xi$ is negative, it can be observed from
(\ref{eq4: umbral xi - pi}) that the threshold energy for
photo-pion production can be easily shifted, preventing the
reaction to take place for high energies. The condition for this
to be the case is
\begin{eqnarray}
- \xi &>& \frac{128}{27} \frac{\omega_{0}^{3}}{m_{\pi}^{3} m_{p}}
\left( \frac{ m_{p} +  m_{\pi} }{ 2 m_{p} + m_{\pi} } \right)^{2}
[\omega / \omega_{0}]^{3} \nonumber\\
&& =  7.75 \times 10^{-45}  \, [\omega / \omega_{0}]^{3} \,\,
\mathrm{eV}^{-1}.
\end{eqnarray}
It can be seen therefore that in the particular case of $\xi = -
l_{p}$ ($l_{p} = 8.3 \times 10^{-29}$ eV$^{-1}$), the reaction can
be considerably suppressed.

Let us note, however, that the pair production $p + \gamma
\rightarrow p + e^{+} + e^{-}$ imposes strong restrictions over a
negative $\xi$ parameter. Following the methods that we have used
up to now, it is possible to find that the threshold condition for
the pair production is
\begin{eqnarray}
\omega E + \xi \frac{m_{e} m_{p}}{(m_{p} + 2 m_{e})^{2}} E^{3}
\geqslant m_{e} (m_{p} + m_{e}).
\end{eqnarray}
Since we cannot infer modifications in the description of pair
production in the cosmic ray spectra up to energies $E \sim 4
\times 10^{19}$ eV, we must at least impute the following
inequality
\begin{eqnarray} \label{eq: bound xi} | \xi |  & < & \frac{(m_{p} + m_{e}) (m_{p} + 2
m_{e})^{2}}{m_{p} \, E_{\mathrm{ref}}^{3}} \nonumber \\ && = 3.26
\times 10^{-41} \,\, \mathrm{eV}^{-1}
\nonumber \\
&& (= 3.98 \times 10^{-13} \, l_{p} ),
\end{eqnarray}
where we have used $E_{\mathrm{ref}} = 3 \times 10^{19}$ eV. This
last result shows the strong suppression over $\xi \,$. As a
consequence, the particular case $|\xi| = l_{p} = 8 \times
10^{-28}$ eV$^{-1}$ should be discarded. A bound like (\ref{eq:
bound xi}) seems to have been omitted up to now for the GZK
anomaly analysis.


\section{Modified Spectrum} \label{MOD-SPEC}

In this section we shall show how the only surviving LQG
correction from our previous analysis in section \ref{Trh},
$A=1+\alpha$, can affect the prediction of the theoretical cosmic
ray spectrum. Our approach will be centered on the supposition
that the LQG corrections to the main quantities for the
calculation ---such as cross sections and inelasticities of
processes--- are, in a first instance, kinematical corrections,
and that the Lorentz symmetry is spontaneously broken. These
assumptions will allow us to introduce the adequate corrections
when a modified dispersion relation is known.

\subsection{Kinematics}

When having a spontaneous Lorentz symmetry breaking, we can use
the still valid Lorentz transformations to express physical
quantities observed in one reference system, in another one. This
is possible since, under a spontaneous symmetry breaking, the
group representations of the broken group preserves its
transformation properties. In particular, it will be possible to
relate the observed 4-momenta in different reference systems
through the usual rule
\begin{eqnarray} \label{eq4: E'-E,p}
p_{\mu}' = \Lambda_{\mu}^{\,\,\, \nu} p_{\nu},
\end{eqnarray}
where $p_{\mu} = (-E, \vec{p})$ is an arbitrary 4-momentum
expressed in a given reference system $\mathcal{K}$, $p_{\mu}'$ is
the same vector expressed in another given system $\mathcal{K}'$,
and $\Lambda_{\mu}^{\,\,\, \nu}$ is the usual Lorentz
transformation connecting both systems. Such a transformation will
keep invariant the scalar product
\begin{eqnarray} \label{eq4: prod p-p}
p^{\mu} p_{\mu} = - E^{2} + p^{2},
\end{eqnarray}
as well as any other product.

Let us illustrate, for transformation (\ref{eq4: E'-E,p}), the
situation in which $\mathcal{K}'$ is a reference system with the
same orientation of $\mathcal{K}$ and which represents an observer
with velocity $\vec \beta$ with respect to $\mathcal{K}$. In this
case $\Lambda_{\mu}^{\,\,\, \nu}$ shall correspond to a boost in
the $\hat \beta = \vec \beta /|\beta|$ direction, and expression
(\ref{eq4: E'-E,p}) will be reduced to
\begin{eqnarray}
E' &=& \gamma (E - \vec \beta \cdot \vec p), \label{eq4: E prima}\\
\vec p \,\, ' &=& \gamma (\vec p - \vec \beta E), \label{eq4: p
prima}
\end{eqnarray}
where $\gamma = (1 - \beta^{2})^{-1/2}$. A particular case of this
transformation will be that in which $\vec \beta$ has the same
direction as $\vec p$, and $\mathcal{K}'$ corresponds to the c.m.
reference system, that is to say, the system in which $\vec p \,\,
' = 0$. In such a case we will have $\vec \beta = \vec p /E$ and
$\gamma = E/(E^{2} - p^{2})^{-1/2}$, jointly with the relation
\begin{eqnarray} \label{eq4: E' - gamma}
E' = E/\gamma = (E^{2} - p^{2})^{1/2}.
\end{eqnarray}
In other words, the c.m. energy of a particle with energy $E$ and
momentum $\vec p$ in $\mathcal{K}$ will correspond to the
invariant $(E^{2} - p^{2})^{1/2}$. Furthermore, such energy is the
minimum measurable energy by an arbitrary observer; this can be
confirmed by solving equation $\partial E' / \partial \beta = 0$
from the relation (\ref{eq4: E prima}) and by verifying that the
solution is $\beta = p/E$. This allows us to interpret $E' =
(E^{2} - p^{2})^{1/2}$ as the rest energy of the given particle.
To simplify the notation and the ensuing discussions, let us
introduce the variable $s = (E')^{2}$, where $E'$ is given by
(\ref{eq4: E' - gamma}).

So far in our analysis, relativistic kinematics has not been
modified. Nevertheless, a difference with the conventional
kinematical frame is that in the present theory the product
(\ref{eq4: prod p-p}) will not be independent of the particle's
energy; conversely, we will have the general expression
\begin{eqnarray} \label{eq4: pp - f}
p^{\mu} p_{\mu} =  - f_{a}(E,\vec{p}) - m_{a}^{2},
\end{eqnarray}
where $f_{a}(E,\vec{p})$ is a function of the energy and the
momentum, that represents the LIV provided by the LQG effective
theories. Let us note that expression (\ref{eq4: pp - f}) is just
the modified dispersion relation
\begin{eqnarray} \label{eq4: E - p f}
E^{2} =  p^{2} + f_{a}(E,\vec{p}) + m_{a}^{2}.
\end{eqnarray}
To be consistent $f_{a}(E,\vec{p})$ must be invariant under
Lorentz transformations and, therefore, can be written as a scalar
function of the energy and the momentum.

We have already made mention of the fact that LIV's inevitably
introduce the appearance of a privileged system; in the present
discussion we will choose as such a system the isotropic system
(which by assumption is the co-moving CMBR system), and will
express $f_{a}(E,\vec{p})$ in terms of $E$ and $\vec{p}$ measured
in that system. As may be expected in this situation, $f_{a}$ will
be a function only of the energy $E$ and the momentum norm
$p=|\vec{p}|$, since no trace of a vectorial field could be
allowed when isotropy is imposed. For example, in the particular
case of the dispersion relation for a fermion, the function
$f_{a}(E,\vec{p})$ depends uniquely on the momentum, and can be
written as
\begin{eqnarray}
f_{a}(p) = 2 \alpha_{a} p^{2} + \eta_{a} p^{4} \pm 2 \lambda_{a}
p.
\end{eqnarray}
For simplicity, we shall continue using $f_{a}(p)$ instead of
$f_{a}(E,\vec{p})$.

Through the recently introduced notation and the use of expression
(\ref{eq4: E' - gamma}), the c.m. energy of an $a$ particle with
mass $m_{a}$ and deformation $f_{a}(p)$ will be
\begin{eqnarray} \label{eq4: s - f}
s_{a}^{1/2} = \sqrt{f_{a}(p) + m_{a}^{2}}.
\end{eqnarray}
Of course, the validity of this interpretation will be subordinate
to those cases in which
\begin{eqnarray} \label{eq4: restric tipo tiempo}
s_{a} = f_{a}(p) + m_{a}^{2} > 0,
\end{eqnarray}
or, equivalent, to those states with a time-like 4-momentum. In
the converse, particles with energies and corrections such that
$s_{a} = f_{a}(p) + m_{a}^{2} \leqslant 0$ will be described by
light-like physical states if the equality holds, or space-like
physical states if the inequality holds.

A new effect provided by LIV's is that, if a reference system
where $p=0$ exists, then in that system the particle will not be
generally at rest. To understand this it is sufficient to verify
that in general the velocity follows
\begin{eqnarray}
v = \frac{\partial E}{\partial p} \neq \frac{p}{E},
\end{eqnarray}
and therefore does not vanish at $p=0$. Returning to equations
(\ref{eq4: E prima}) and (\ref{eq4: p prima}), we can see that
when $\beta = v = \partial E / \partial p $, the following result
is produced
\begin{eqnarray} \label{eq4: v' -- 0}
\frac{\partial E '}{\partial p '} = \gamma_{v} \left(
\frac{\partial E}{\partial p} - v \right) \frac{\partial
p}{\partial p '} = 0,
\end{eqnarray}
where $\gamma_{v} = (1 - v^{2})^{-1/2}$. Result (\ref{eq4: v' --
0}) shows that the velocity of the system where the particle is at
rest is effectively $v$. There emerges, then, an important
distinction between the phase velocity $\beta = p/E$ of the c.m.
system of a particle, and the group velocity $v =
\partial E / \partial p$ of the same particle.

The above results, for a single particle, can be easily
generalized to a system of many particles. For instance, the total
4-momentum of a system of many particles $p_{\mathrm{tot}}^{\,
\mu} = \sum_{i} p_{i}^{\, \mu}$ will transform through the rule
(\ref{eq4: E'-E,p}), and the scalar product $(p^{\, \mu}
p_{\mu})_{\mathrm{tot}}$ will be an invariant under Lorentz
transformations (as well as any other product). As in the case of
individual particles, we define $\sqrt{s}$ as the total rest
energy measured in the system of c.m. That is to say:
\begin{eqnarray}
s = E_{\mathrm{tot}}^{2} - p_{\mathrm{tot}}^{2}.
\end{eqnarray}
In the case in which we have a system composed of a proton with
energy $E$ and momentum $p$, and a photon (from the CMBR) with
energy $\omega$ and momentum $k$ (all these quantities are
measured in the laboratory isotropic system $\mathcal{K}$), the
$s$ quantity shall acquire the form
\begin{eqnarray}
s &=& (E + \omega)^{2} - (\vec p + \vec k)^{2} \nonumber \\
  &=& (E' + \omega')^{2} - (\vec p \, ' + \vec k)^{2},
\end{eqnarray}
where the $E'$ and $p'$ quantities are measured in an arbitrary
reference system. In particular we are interested in the system
where the proton momentum $\vec p \,'$ is null; that is to say,
the system in which $E' = \sqrt{s_{p}} \,$. If $\epsilon$ is the
photon energy in such a system, then
\begin{eqnarray}
s &=& (\sqrt{s_{p}} + \epsilon)^{2} - \epsilon^{2} \nonumber \\
  &=& 2 \sqrt{s_{p}} \,\, \epsilon + s_{p} \, ,
\end{eqnarray}
where we have used the dispersion relation $\omega = k$ (or
$\epsilon = k \, '$) for the CMBR photons. These results will
allow us to express the main kinematical quantities in term of
$\epsilon$ and $s_{p} = f_{p}(p) + m_{p}^{2} \,$. For example, the
Lorentz factor that connects the $\mathcal{K}$ system with the
$\mathcal{K}'$ system where $p \, ' = 0$, will be
\begin{eqnarray}
\gamma = \frac{E}{\sqrt{s_{p}}} \, .
\end{eqnarray}
Meanwhile, the Lorentz factor connecting $\mathcal{K}$ with the
c.m. system (that in which $\vec p \, ' + \vec k = 0 $), will be
\begin{eqnarray}
\gamma_{c} &=& \frac{E + \omega}{s_{p} + 2 \sqrt{s_{p}} \,\,
\epsilon} \nonumber\\
&\simeq& \frac{E}{s_{p} + 2 \sqrt{s_{p}} \,\, \epsilon}.
\end{eqnarray}

As a last comment, let us note that to the first order in the
expansion of the dispersion relations in terms of the scales
$\mathcal{L}$ and $l_{p}$, when we consider high energy processes
such that $p^{2} \gg f(p) + m^{2}$ we can freely interchange the
momentum $p$ by the energy $E$ in the deviation function $f(p)$.
That is to say, we may consider as a valid relation the following
expression
\begin{eqnarray}
E^{2} = p^{2} + f(E) + p^{2},
\end{eqnarray}
where we have made the replacement $f(p) \rightarrow f(E)$. This
procedure will greatly simplify the next discussions.

\subsection{Modified Inelasticity: $p+\gamma \rightarrow p+x$}

Following the same methods of section \ref{UHECR}, let us obtain a
modified inelasticity $K$ for a process of the type $p+\gamma
\rightarrow p+x$, where $x$ is an emitted particle that, in the
present physical problem in which we are interested, can be a
$\pi$, $\rho$ or $\omega$ meson. We note that the dispersion
relation for the emerging proton (after a collision with a photon)
can be written in the form:
\begin{eqnarray} \label{eq5: Kmod: desarrollo}
E_{p}^2-p_{\, p}^2= f_{p}(E_{p})+m_{p}^{2},
\end{eqnarray}
where $E_{p}$ is the final proton energy. Since the left side of
(\ref{eq5: Kmod: desarrollo}) is invariant under Lorentz
transformations, we can write
\begin{eqnarray}
(E_{p}^{*})^2-(p_{\, p}^{*})^2= f_{p}(E_{p})+m_{p}^{2},
\end{eqnarray}
where the * denotes the quantities measured in the c.m. system. On
the other hand, in such a system, the following conservation
relations of energy and momentum are satisfied:
\begin{eqnarray}
E_{p}^*+E_{x}^*=\sqrt{s}
\end{eqnarray}
and
\begin{eqnarray}
(p_{p}^{*})^2=(p_{x}^{*})^2.
\end{eqnarray}
Substituting both quantities in relation (\ref{eq5: Kmod:
desarrollo}), we can obtain
\begin{eqnarray} \label{eq5: Kmod mas des.}
2 \sqrt{s} E_{p}^*= s + f_{p}(E_{p})-f_{x}(E_{x}) +
m_{p}^2-m_{x}^2,
\end{eqnarray}
or, in a more convenient form
\begin{eqnarray} \label{eq5: Kmod mas des.}
2 \sqrt{s} E_{p}^*= s + s_{p}(E_{p})-s_{x}(E_{x}).
\end{eqnarray}
In the same way, we also have the energy conservation relation in
the laboratory system:
\begin{eqnarray} \label{eq5: cons de E p-x}
E_{p}+E_{x}=E_{\mathrm{tot}}.
\end{eqnarray}
Using the definition for the inelasticity $K_{x} = \Delta E /E$
for a process, where $\Delta E = E_{i}-E_{f} \simeq
E_{\mathrm{tot}} - E_{f}$, it is possible to rewrite (\ref{eq5:
cons de E p-x}) in terms of $K_{x}$ through expressions
\begin{eqnarray}
E_{x} &=& K_{x} E , \\
E_{p} &=& (1 - K_{x}) E,
\end{eqnarray}
where $E$ is the initial energy of the initial proton. Having done
this, equation (\ref{eq5: Kmod mas des.}) now acquires the form:
\begin{eqnarray}
2 \sqrt{s} E_{p}^*= s + s_{p}[(1 - K_{x}) E]-s_{x}[K_{x} E].
\end{eqnarray}
To simplify the development of the inelasticity, let us write the
former relation as $E_{p}^{*}=F(E, K_{x})$, where $F = F(E,K_{x})$
is defined through
\begin{eqnarray} \label{eq5: def F}
F = \frac{1}{2\sqrt{s}}( s + s_{p}[(1 - K_{x}) E]  -s_{x}[K_{x}
E]).
\end{eqnarray}
On the other side, the Lorentz transformation rules give us the
relation between the proton energies in the laboratory system and
the c.m. system respectively. This relation is
\begin{eqnarray}
E_{p} &=& \gamma_{c} (E_{p}^{*} + \beta_{c} p_{p}^{*} \cos \theta)
\nonumber\\
&=& \gamma_{c} \left( E_{p}^{*} + \beta_{c} \sqrt{E_{p}^{*2} -
s_{p}(E_{p})} \cos \theta \right). \label{eq5: Ep}
\end{eqnarray}
Joining (\ref{eq5: def F}) and (\ref{eq5: Ep}), it is possible to
find the general equation for $K_{x}$:
\begin{eqnarray} \label{eq5: ec para Kx}
(1-K_{x}) \sqrt{s} =  \Big( F(E,K_{x}) \qquad \qquad \qquad \qquad \nonumber\\
 + \sqrt{F^{2}(E,K_{x}) - s_{p}[(1-K_{x})E]} \cos \theta \Big).
\end{eqnarray}
It should be noted, however, that the solution for $K_{x}$ from
(\ref{eq5: ec para Kx}) will depend in the $\theta$ angle. For
this reason, once this last equation is resolved, it is convenient
to define the total inelasticity $K$ as the average of $K_{x}$
with respect to the $\theta$ angle. That is to say
\begin{eqnarray} \label{eq5: K mod}
K = \frac{1}{\pi} \int_{0}^{\pi} K_{x} \, d \theta.
\end{eqnarray}
It is relevant to mention that now, as opposed to result
(\ref{eq2: K}), the inelasticity $K$ will be a function of both,
the energy $E$ of the initial proton and the energy $\epsilon$ of
the CMBR photon. 

\subsection{The $m_{a}^{2} \rightarrow s_{a} = m_{a}^{2} + f_{a}(E)$
Prescription} \label{sec: prescription}

Let us recall our interpretation relative to the fact that
$s_{a}^{1/2} = (f_{a}(E_{a}) + m_{a}^{2})^{1/2}$ can be understood
as the rest energy of a particle $a$, as a function of the energy
$E_{a}$ that it has in the laboratory system $\mathcal{K}$. As we
have already emphasized, such an interpretation will be valid for
particles with time like 4-momenta.

In the reactions given between high energy protons and the photons
of the CMBR, the whole scenario consists of the collision between
two particles, $p$ and $\gamma$, with the subsequent production of
a certain number of final particles. Let us suppose that $a$ is
one of these particles in the final state. The knowledge of the
inelasticity $K$ for the reaction will allow us to estimate the
average energy $<E_{a}>$ with which such a particle emerges (since
$K$ provides the average fraction of energy with which such a
particle is produced). That is to say, on average, the rest energy
of the final particle $a$ will be $s_{a}^{1/2} = [f_{a}(<E_{a}>) +
m_{a}^{2}]^{1/2}$. Moreover, the knowledge of the inelasticity $K$
will allow us to express $s_{a}$ as a function of the energy $E$
of the initial proton:
\begin{eqnarray} \label{eq5: sa - sa(E)}
s_{a} = s_{a}(E).
\end{eqnarray}
Following our previous interpretation, we can view the recently
described process as a reaction between a proton with mass
$s_{p}^{1/2}$, which loses energy emitting particles $a$ with mass
$s_{a}^{1/2}$ calculated in the previous form. This idealized
reasoning gives us a clear prescription to kinematically modify
those dynamical quantities with which we must work and where
energy conservation is involved. This prescription is:
\begin{eqnarray} \label{eq5: prescrip}
m_{a}^{2} \rightarrow s_{a}(E) = f_{a}(E) + m_{a}^{2},
\end{eqnarray}
where we have expressed correction $f_{a}$ as a function of the
initial energy of the incident proton.

Prescription (\ref{eq5: prescrip}) establishes the notion of an
effective mass which is dependent on the initial energetic content
of a reaction. As a consequence, given the explicit knowledge of the
dependence that a cross section has on the masses and energies of
the involved states, to obtain the modified version, it will be
appropriate to use the discussed prescription.

Let us note, however, that an important weakness of the present
method is the inability to determine whether $m_{p}$ comes from
the initial state proton (with an initial energy $E$), or the
final state proton (with a final energy $E_{p} = (1 - K) E$);
specially since that distinction gives rise to different values
for $s_{p}$. To overcome this difficulty, and therefore any
ambiguity in the prescription, we shall restrict our treatment to
the case $\alpha_{p} = 0$.

\subsection{Redshift}

Another important problem related to the introduction of LIV's in
the dispersion relations is whether the redshift relation for the
propagation of particles in a FRW universe is modified. This could
be of great relevance because of the large distances involved in
cosmic ray propagation and, therefore, the possible cumulative
effects. We shall examine this issue through the study of a
classical point-like particle propagating in a FRW space-time.

The most general action for a point particle in a given space-time
is
\begin{eqnarray}
S = \int_{a}^{b} \Lambda \, d\tau,
\end{eqnarray}
with $\tau$ an affine parameter chosen to accomplish $d \tau =
\sqrt{-g_{\mu \nu}dx^{\mu}dx^{\nu}}$ (we are using the mostly plus
signature $(-,+,+,+)$). The variation of the action can be
realized through two separated terms:
\begin{eqnarray} \label{eq: delta-s}
\delta S =  \int_{a}^{b} \Lambda \, \delta d\tau + \int_{a}^{b}
\delta \Lambda \, d\tau.
\end{eqnarray}
Starting with the first term, it is possible to write $\delta d
\tau$ in the following way (using $u^{\mu} = d x^{\mu} / d \tau$
and identifying the free torsion connections $\Gamma_{\mu
\sigma}^{\nu}$):
\begin{eqnarray}
\delta d \tau &=& -g_{\mu \nu} \frac{dx^{\mu}}{d \tau} \delta
dx^{\nu} - \Gamma_{\mu \sigma}^{\nu} u^{\mu} u_{\nu} \, d \tau \,
\delta x^{\sigma}. \label{delta-tau}
\end{eqnarray}
Naturally, in the above expression we can allow $\delta dx^{\mu}=
d (\delta x^{\mu})$ and $\delta d \tau = d (\delta \tau)$. The
variation of the second term in (\ref{eq: delta-s}) can be
developed through
\begin{eqnarray}
\delta \Lambda = \frac{\partial \Lambda}{\partial x^{\sigma}}
\delta x^{\sigma} + \frac{\partial \Lambda}{\partial u^{\sigma}}
\delta u^{\sigma}.
\end{eqnarray}
For the $\delta u^{\mu}$ variation we must proceed carefully since
there will be constraints between $\delta u^{\mu}$ and $\delta
x^{\mu}$. We have $\delta u^{\mu}= u'^{\mu} - u^{\mu}$, where
$u'^{\mu} = dx'^{\mu} / d \tau' $, with $dx'^{\mu}
=dx^{\mu}+\delta x^{\mu}$ and $d \tau' = d \tau + \delta \tau$. In
this way, to the first order in the variations, it is possible to
find the following constraint between $\delta u^{\mu}$ and $\delta
x^{\mu}$ (where we have used equation (\ref{delta-tau}) to work
out the relation):
\begin{eqnarray}
(\delta u^{\mu}) d\tau &=& d(\delta x^{\mu}) + u^{\mu} [ g_{\eta
\nu} u^{\eta} \delta dx^{\nu} \nonumber\\
&& + \Gamma_{\mu \sigma}^{\nu} u_{\nu} \, d \tau \, \delta
x^{\sigma} ].
\end{eqnarray}
Using the former relations, the complete expression for the
variation of the action is
\begin{eqnarray} \label{eq5: var accion}
\delta S  = \left[  \frac{\partial \Lambda}{\partial u^{\mu}} +
\Delta u_{\mu} \right] \delta x^{\mu} \bigg |_{a}^{b} +
\int_{a}^{b} \bigg[   \frac{\partial \Lambda}{\partial x^{\mu}} \qquad \nonumber\\
 + \Delta \Gamma_{\mu \nu}^{\eta}
 u_{\eta} u^{\nu} - \frac{d}{d \tau} \left( \frac{\partial \Lambda}{\partial
u^{\mu}} + \Delta u_{\mu} \right) \bigg] d \tau \delta x^{\mu},
\end{eqnarray}
where we have defined $\Delta \equiv (\partial \Lambda / \partial
u^{\sigma})u^{\sigma}-\Lambda$.  Let us now define the momenta
$p_{\mu}$ as follows
\begin{eqnarray}
p_{\mu} \equiv  \frac{\delta S}{\delta x^{\mu}} \bigg |_{b} =
\frac{\partial \Lambda}{\partial u^{\mu}} + \Delta u_{\mu}.
\label{eq:p}
\end{eqnarray}
This definition is concomitant with the canonical approach and
allow us to write the equations of motion in a very simple and
convenient way:
\begin{eqnarray} \label{eq:dp}
dp_{\mu}= \left[ \frac{\partial \Lambda}{\partial x^{\mu}}+ \Delta
\Gamma^{\eta}_{\mu \nu} u_{\eta} u^{\nu} \right] d \tau.
\end{eqnarray}

We are interested in obtaining an expression for the redshift
relation having as a starting point the equations of motion
(\ref{eq:dp}) deduced from the $S$ action. For this we must
consider the FRW metric
\begin{eqnarray}
g_{\mu \nu} = \mathrm{diag} \left[ -1,\frac{R^{2}(t)}{1-k
r^{2}},R^{2}(t)r^{2},R^{2}(t)r^{2} \sin^{2}{\theta} \right].
\end{eqnarray}
To accomplish our goal, let us calculate the variation of
$p^{2}=g^{ij}p_{i}p_{j}$ through a path parameterized by $\tau$:
\begin{eqnarray} \label{eq:p^2}
\frac{d}{d \tau}p^{2} &=& \frac{d}{d \tau} (g^{ij}p_{i}p_{j}) \nonumber\\
&=& ( \frac{d}{d \tau}g^{ij} ) p_{i}p_{j} + 2g^{ij}p_{i}
\frac{dp_{j}}{d \tau}.
\end{eqnarray}
Note that inn the preceding expression, $dp_{j}$ is given by the
dynamical equations (\ref{eq:dp}). Using these equations to
simplify equation (\ref{eq:p^2}), it is possible to deduce that
\begin{eqnarray}
\frac{d}{d \tau}(p R) =  \frac{R}{p} g^{ij} p_{i} \, \Omega_{j},
\end{eqnarray}
where $\Omega_{j}$ is defined through
\begin{eqnarray}
\Omega_{i} \equiv \left( \frac{\partial \Lambda}{\partial x^{i}} -
u^{j} \Gamma^{k}_{i j} \frac{\partial \Lambda}{\partial u^{k}}
\right),
\end{eqnarray}
and the $\Gamma^{k}_{ij}$ are the FRW spatial connections given by
\begin{eqnarray}
\Gamma^{k}_{ij} &=& \frac{1}{2}g^{kl}(g_{il,j}+g_{jl,i}-g_{ij,l}).
\label{con1}
\end{eqnarray}
To conclude, we now turn our attention to $\Omega_{i}$. If we want
to be loyal to the spirit of the FRW space-time formulation, we
must impose isotropy and homogeneity on the Lagrangian $\Lambda$,
when expressed in the co-moving FRW frame (a characteristic
present in our previous development of LIV's). That is to say,
$\Lambda = \Lambda(u^2, t)$, where $u^2=g_{ij}u^{i}u^{j}$. In this
way, the spatial dependence will be through $g_{ij}$. If we
differentiate $\Lambda$ with respect to $x^{k}$ then
\begin{eqnarray}
\frac{\partial \Lambda}{\partial x^{k}} &=& \frac{\partial
\Lambda}{\partial u^{2}} \frac{\partial u^{2}}{\partial x^{k}} \nonumber\\
 &=& \frac{\partial \Lambda}{\partial u^{2}} u^{i}u^{j}
\frac{\partial g_{ij}}{\partial x^{k}},
\end{eqnarray}
which implies that
\begin{eqnarray}
\frac{\partial \Lambda}{\partial x^{k}} &=& \frac{\partial
\Lambda}{\partial u^{i}} u^{j} \Gamma^{i}_{jk}.
\end{eqnarray}
Note that this in turn will mean that $\Omega_{i}=0$. So, as a
general result, the usual redshift relation is reobtained:
\begin{eqnarray}
\frac{d}{d \tau}(p R) =  0.
\end{eqnarray}

\subsection{Spectrum and Results}

Introducing the above modifications to the different quantities
involved in the propagation of protons (like the cross section
$\sigma$ and inelasticity $K$), we are able to find a modified
version for the UHECR energy loss due to collisions. Since the
only relevant correction for the GZK anomaly is $\alpha$, we
focused our analysis on the particular case $f(p) = 2 \alpha
p^{2}$. To simplify our model we restricted our treatment to the
case $\alpha > 0$ (consistent with the effective mass
interpretation) and used only $\alpha_{m} \neq 0$, where
$\alpha_{m}$ is assumed to have the same  value for mesons $\pi$,
$\rho$ and $\omega$.

\begin{figure}[ht] 
\begin{center}
\includegraphics[width=0.48\textwidth]{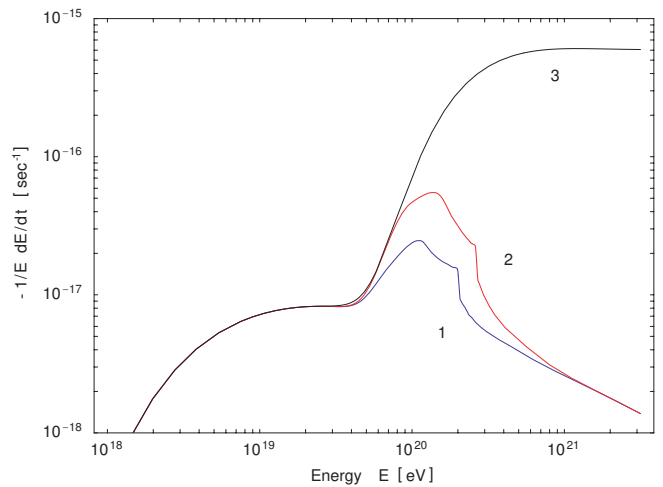}
\caption[Modified energy loss]{Modified energy loss for UHECR due
to collisions. The figure shows the case $\alpha_{m} \neq 0$, for
three different values of the weave scale $\mathcal{L}$. Curve 1:
$\alpha_{m} = 9 \times 10^{-23}$ ($\mathcal{L} \simeq 8.6 \times
10^{-18}$ eV$^{-1}$); curve 2: $\alpha_{m} = 5 \times 10^{-23}$
($\mathcal{L} \simeq 1.2 \times 10^{-17}$ eV$^{-1}$); curve 3:
$\alpha_{m} = 0$ (without modifications).} \label{F3}
\end{center}
\end{figure}

Fig.\ref{F3} shows the modified energy loss $\tau (E)$ for UHECR
obtained for different values of $\alpha_{m}$. These are, curve 1:
$\alpha_{m} = 9 \times 10^{-23}$ ($\mathcal{L} \simeq 8.6 \times
10^{-18}$ eV$^{-1}$); curve 2: $\alpha_{m} = 5 \times 10^{-23}$
($\mathcal{L} \simeq 1.2 \times 10^{-17}$ eV$^{-1}$); and curve 3:
$\alpha_{m} = 0$ which corresponds to the case without
modifications given by the conventional theory.
It can be seen therefore how the corrections can affect the main
life time of protons propagating through the CMBR, allowing  a
strong improvement in the distances that protons can reach before
loosing their characteristic energy (for energies greater than $1
\times 10^{20}$ eV). The effects that the LQG corrections
have on the propagation of UHECR are manifest through a decay of
the energy loss in the range $E \sim 1 \times 10^{20}$ eV. To
understand this, recall relation (\ref{eq: pion2}) for the
threshold condition of photo-pion production:
\begin{eqnarray}
2 \, \delta \alpha \, E_{\pi}^{2} + 4E_{\pi} \omega \geqslant
\frac{m_{\pi}^{2}(2 m_{p}+m_{\pi})}{m_{p}+m_{\pi}}.
\end{eqnarray}
As we saw in section \ref{Trh}, the condition for a significant
increase or decrease in the energy threshold can be calculated as
$|\delta \alpha| \geqslant (2 m_{p} + m_{\pi})(m_{p} + m_{\pi})/2
E^{2}$. Therefore, for a given value of $\delta \alpha > 0$, the
energy at which the LIV effects start to take place is
\begin{eqnarray}
E^{2} = \frac{1}{2 \delta \alpha} (2 m_{p} + m_{\pi})(m_{p} +
m_{\pi}).
\end{eqnarray}
In the case $\alpha_{m} = 9 \times 10^{-23}$ (curve 1 of
Fig.\ref{F3}), this energy is $E=1.1 \times 10^{20}$ eV, while in
the case $\alpha_{m} = 5 \times 10^{-23}$ (curve 2) this
corresponds to $E=1.5 \times 10^{20}$ eV. Beyond these energy
scales, at about $E \sim 2 \times 10^{20}$ eV, a sharp decay is
observed in the behavior of the curve. This is due to the fact
that the modified inelasticity $K$ will strongly constraint the
energy-momentum phase space accessible to the final states
depending on the initial energy $E$ that the primary proton carries
(recall that now $K$ is a function of the energy $E$ of the
incident proton and the energy $\epsilon$ of the CMBR photon).

\begin{figure}[ht] 
\begin{center}
\includegraphics[width=0.48\textwidth]{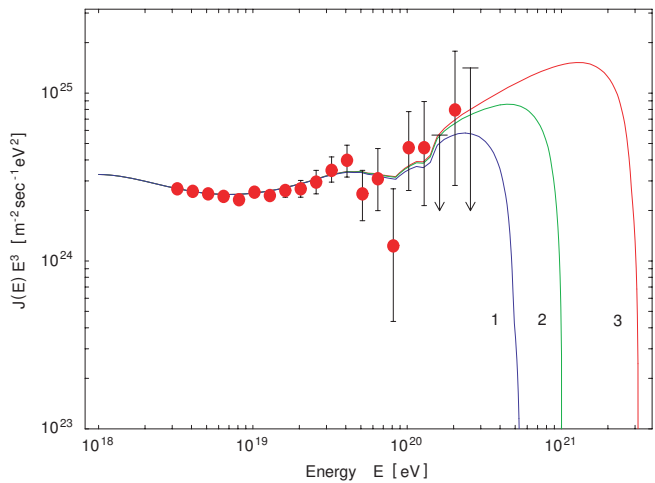}
\caption[Modified cosmic ray spectrum]{Modified UHECR spectrum and
AGASA observations. The figure shows the modified spectrum $J(E)$
multiplied by $E^{3}$, for uniform distributed sources and without
evolution, for the case $\alpha_{m} = 1.5 \times 10^{-22}$
($\mathcal{L} \simeq 6.7 \times 10^{-18}$ eV$^{-1}$). Three
different maximum generation energies $E_{\mathrm{max}}$ are
shown. These are, curve 1: $5 \times 10^{20}$ eV; curve 2: $1
\times 10^{21}$ eV; and curve 3: $3 \times 10^{21}$ eV.}
\label{F4}
\end{center}
\end{figure}

We can also find the modified version of the UHECR spectrum for
$\alpha_{m} \neq 0$. Fig.\ref{F4} shows the AGASA observations and
the predicted UHECR spectrum in the case $\alpha_{m} = 1.5 \times
10^{-22}$ ($\mathcal{L} \simeq 6.7 \times 10^{-18}$ eV$^{-1}$) for
three different maximum generation energies $E_{\mathrm{max}}$.
These are, curve 1: $5 \times 10^{20}$ eV; curve 2: $1 \times
10^{21}$ eV; and curve 3: $3 \times 10^{21}$ eV.
The Poisson probabilities of an excess in the five highest energy
bins for the three curves are $P_{1} = 3.6 \times 10^{-4}$, $P_{2}
= 2.6 \times 10^{-4}$ and $P_{3} = 2.3 \times 10^{-4}$. The
Poisson $\chi^{2}$ for the eight highest energy bins are
$\chi^{\,\, 2}_{1} = 10$, $\chi^{\,\, 2}_{2} = 10.9$ and
$\chi^{\,\, 2}_{3} = 11.2$ respectively. The possibility of
reconciling the data with finite maximum generation energies is
significant given that conventional models require infinite
maximum generation energies $E_{\mathrm{max}}$ for the best fit.
 For the lower part of the spectrum (under $E = 4 \times
10^{19}$ eV), the parameters under consideration leave the
spectrum completely unaffected. This is due to the fact that in
such a region the dominant reaction is the pair production, which
has not being modified to obtain the spectrum. A more accurate
study on this issue would require the computation of a modified
inelasticity for the pair creation. Meanwhile, we must content
ourself with the semiqualitative criteria given in section
\ref{Trh} to rule out the parameters.

\section{Conclusions} \label{CONC}

The scientific challenge that represents the search for new
empirical backgrounds to test quantum gravity theories is at the
embryonic stage. In this context, the possibility that ultra high
energy cosmic rays could be experiencing quantum gravity effects
places us in a very challenging situation which deserve attention.
Nevertheless, the present stage of UHECR observations demands that
we proceed with caution and patience.

We have seen how the kinematical analysis of the different
reaction taking place in the propagation of ultra high energy
protons can set strong bounds on the parameters to the theory. In
comparison with our previous work, we have eliminated some
previously open possibilities by the particular study of the pair
creation $p + \gamma \rightarrow p + e^{+} + e^{-}$, in the energy
region where this reaction dominates the proton's interactions
with the CMBR. In this way, the only possibility still open (for
the corrective terms considered in the expansion (\ref{eq3: E for
fermions general}) for the dispersion relations) and favored by
the LQG scales, is the correction $\alpha$. If this is the case, a
favored region for the scale length $\mathcal{L}$ estimated
through the threshold analysis would be
\begin{eqnarray}
2.6 \times 10^{-18} \,\,\, \mathrm{eV}^{-1} \lesssim \mathcal{L}
\lesssim 1.6 \times 10^{-17} \,\,\, \mathrm{eV}^{-1}. \nonumber
\end{eqnarray}

Similarly, the kinematical corrections can be studied in more
detail when their effects are considered in the theoretical
spectrum. In this regard, we have seen how to develop a modified
version of the inelasticity for the photo-pion production, and its
implications in the mean life time of a high energy proton as well
as on the spectrum. To accomplish this last task we have only
assumed a spontaneous Lorentz symmetry break up in the effective
equations of motion, allowing the use of Lorentz transformations
on the dispersion relations. Therefore, result (\ref{eq5: ec para
Kx}) can be used in a more general context than the special case
offered by the LQG framework.

Special mention must be made of a recent development \cite{AMU},
where the dispersion relations for fermions and bosons are
generalized to include the extra factor $\Upsilon$. It should be
noted that this new factor always appears in the dispersion
relations in the form $\left( l_{p} / \mathcal{L}
\right)^\Upsilon$, such that the parameter $\alpha$ in our
equation (\ref{eq: E for fermions}) gets an extra factor $\left(
l_{p} / \mathcal{L} \right)^{\Upsilon-1}$. This freedom can be
used to move the scale $\cal L$ down if needed, so that the cosmic
ray momentum $p$ always satisfies the bound $p \, \mathcal{L}
\leqslant 1$, without changing our prediction of the UHECR
spectrum.

Future experimental developments like the Auger array, the Extreme
Universe Space Observatory (EUSO) and Orbiting Wide-Angle Light
Collectors (OWL) satellite detectors, will increase the precision
and phenomenological description of UHECR. On the more theoretical
side, progress in the direction of a full effective theory, with a
systematic method to compute any correction with a known value for
each coefficient, is one of the next steps in the ``loop''
quantization programme \cite{Thiemann2, Thiemann3}. Therefore, it
is important to trace a phenomenological understanding of the
possible effects that could arise as well as the constraints on
LQG, in the high and low energy regimens (for other
phenomenological studies of LQG effects, see for example
\cite{SUV} and \cite{Lambiase}).

\section*{Acknowledgements}
We thank J. Ellis for a useful discussion, and D.R. Bergman for
the HiRes data. The work of J.A. is partially supported by
Fondecyt 1010967. He acknowledges the hospitality of LPTENS
(Paris) and CERN; and financial support from an
Ecos(France)-Conicyt(Chile) project. The work of G.P. is partially
supported by a CONICYT Fellowship.


\end{document}